\documentclass[11pt]{article}
\usepackage[USenglish]{babel}
\usepackage{a4wide}
\usepackage{epsfig}
\usepackage[all]{xy}
\usepackage{amssymb}
\usepackage{amsmath}
\usepackage{latexsym}
\usepackage{pifont}
\usepackage{pdfsync}
\usepackage{enumerate}
\usepackage[applemac]{inputenc} 
\title{Towards a Coalgebraic Interpretation of Propositional Dynamic Logic\thanks{Research funded in part by Deutsche Forschungsgemeinschaft, grant DO 263/12-1, \emph{Koalgebraische Eigenschaften stochastischer Relationen.}}}
\author{Ernst-Erich Doberkat\\Technische Universitt Dortmund\\\texttt{ernst-erich.doberkat@udo.edu}}
\date{\today}
\newcommand{\labelImpl}[2]{\ensuremath{\ref{#1}~\Rightarrow~\ref{#2}}}

\newcommand{\Klasse}[2]{\left[#1\right]_{#2}}
\newcommand{\Faktor}[2]{{#1}/{#2}}
\newcommand{\fMap}[1]{\eta_{#1}}
\newcommand{\Bild}[2]{{#1}\left[#2\right]}
\newcommand{\InvBild}[2]{\Bild{#1^{-1}}{#2}}
\newcommand{\Kern}[1]{\mathsf{ker}\left(#1\right)}
\newcommand{\Folge}[1]{(#1_n)_{n \in \Nat}}

\newcommand{\theTheory}[2]{Th_{#1}({#2})}
\newcommand{\spaceFont}[1]{\mathfrak{#1}}

\newcommand{\SubProb}[1]{\spaceFont{S}\left(#1\right)}

\newcommand{\SubProbSenza}{\spaceFont{S}}

\newcommand{\Borel}[1]{\ensuremath{{\mathcal B}(#1)}}

\edef\LinkeKlammer{\lbrack\!\lbrack}
\edef\RechteKlammer{\rbrack\!\rbrack}
\newcommand{\Gilt}[1][\phi]{\ensuremath{\LinkeKlammer#1\RechteKlammer}}

\newcommand{\Trans}{\rightsquigarrow}

\newcommand{\compositionSymbol}[1]{\mathbf{#1}}

\newcommand{\klComp}{\compositionSymbol{\ast}}

\newtheorem{definition}{Definition}[section]
\newcommand{\BeginDefinition}[1]{%
  \begin{definition}\label{#1}
}
\newcommand{\EndDefinition}{\end{definition}}

\newtheorem{example}[definition]{Example}
\newcommand{\BeginExample}[1]{%
  \begin{example}\label{#1}\rm
}
\newcommand{\EndExample}{$\diamondsuit$ \end{example}}

\newtheorem{observation}[definition]{Observation}
\newcommand{\BeginObservation}[1]{
  \begin{observation}\label{#1}\rm
}
\newcommand{\EndObservation}{--- \end{observation}}

\newtheorem{theorem}[definition]{Theorem}
\newcommand{\BeginTheorem}[1]{%
  \begin{theorem}\label{#1}
}
\newcommand{\EndTheorem}{\end{theorem}}

\newtheorem{corollary}[definition]{Corollary}
\newcommand{\BeginCorollary}[1]{
  \begin{corollary}\label{#1}
}

\newtheorem{proposition}[definition]{Proposition}
\newcommand{\BeginProposition}[1]{%
  \begin{proposition}\label{#1}
}
\newcommand{\EndProposition}{\end{proposition}}

\newcommand{\EndCorollary}{\end{corollary}}
\newtheorem{lemma}[definition]{Lemma}
\newcommand{\BeginLemma}[1]{%
  \begin{lemma}\label{#1}
}
\newcommand{\EndLemma}{\end{lemma}}

\newtheorem{claim}{Claim}
\newcommand{\BeginClaim}[1]{%
  \begin{claim}\label{#1}
}
\newcommand{\EndClaim}{\end{claim}}

\newenvironment{proof}{\textbf{Proof\ }}{\ensuremath{\QED}}
\newcommand{\BeginProof}{\begin{proof}}
\newcommand{\EndProof}{\end{proof}}

\newenvironment{remark}{\textbf{Remark:\ }}{}
\newcommand{\BeginRemark}{\begin{remark}}
\newcommand{\EndRemark}{\QED\end{remark}}
\newcommand{\QED}{%
\ensuremath{\dashv}
}

\newcommand{\Real}{\mathbb{R}}

\newcommand{\Nat}{\mathbb{N}}
\newcommand{\Rational}{\mathbb{Q}}

\newcommand{\nPfeil}{\stackrel{\bullet}{\rightarrow}}

\newcommand{\Star}[1]{{#1}^{\ast}}

\def\fontSets{\mathcal}
\def\fontmod{\mathfrak}
\def\fontAddr{\mathsf}
\def\fontCats{\mathbf}
\def\fontFunct{\mathfrak}
\def\fontLog{\mathsf}
\def\Ur{\fontSets{U}}
\def\P{\fontSets{P}}
\def\M{\fontCats{M}}
\def\N{\fontCats{N}}
\def\S{\fontCats{S}}

\def\B{\fontFunct{B}}

\def\R{\fontFunct{R}}
\def\U{\fontFunct{U}}

\def\It{\fontSets{J}}
\def\Rational{\ensuremath{\fontFunct{Rat}_{0, 1}}}
\def\RationalS{\ensuremath{\Rational^{\dagger}}}
\def\Q{\Rational}

\def\V{\fontCats{V}}
\def\BQ{\B^{R}}
\def\BR{\BQ}
\def\BS{\B^{\dagger}}
\def\NQ{\N^{R}}

\def\K{\fontmod{K}}
\def\Kr{\fontCats{K}}
\def\Mo{\fontmod{M}}
\def\No{\fontmod{N}}
\newcommand{\GM}[1][\Mo]{\ensuremath{\Gamma_{#1}}}
\def\L{\fontmod{L}}
\def\basB{\mathfrak{b}}
\newcommand{\PL}[1][\Ur, \P]{\ensuremath{\fontLog{L}(#1)}}
\newcommand{\PM}[1][\Ur, \P]{\ensuremath{\fontLog{M}(#1)}}

\renewcommand{\theTheory}[3][\Mo]{Th_{#3}(#1, #2)}
\newcommand{\Pro}[1][\Ur]{\fontSets{P}(#1)}
\newcommand{\ProExt}[1][\Ur]{\fontSets{E}(#1)}
\newcommand{\ProIrr}[1][\Ur]{\fontSets{I}(#1)}
\def\DistrL{(d_l)}
\def\DistrR{(d_r)}
\def\Emb{(d_\epsilon)}
\def\Star{(d^*)}
\def\AssS{(ass_s)}
\def\AssU{(ass_u)}
\def\IdEpsl{(id_l)}
\def\IdEpsr{(id_r)}
\def\Comm{(comm)}
\def\Idemp{(idm)}
\def\Transp{(transp)}
\def\InfDisj{(dis_\infty)}

\newcommand{\cpl}[2]{\ensuremath{\overline{{#1}}{}^{#2}}}
\def\phi{\varphi}
\def\To{\Rightarrow}
\def\gilt#1{\Gilt[#1]_\Mo}
\def\kgilt#1{\Gilt[#1]_\K}
\def\oben{\overrightarrow}
\newcommand{\adr}[1]{\ensuremath\fontAddr{#1}}
\newcommand{\subTree}[2]{\ensuremath{{#1}|_{\adr{#2}}}}
\newcommand{\subst}[3]{\ensuremath{{#1}[#2]_{\adr{a}}}}

\def\infOp{\bigvee}
\def\ord{w}
\def\md#1#2{\ensuremath{\boldsymbol{\lfloor} #1\boldsymbol{\rceil}_{#2}\,}}

\newcommand{\wrd}[1]{\ensuremath{\Omega(#1)}}
\def\klasse#1{\Klasse{#1}{\equiv}}
\def\faktor#1{\Faktor{#1}{\equiv}}
\def\ePDL{\sim} 
\newcommand{\isEquiv}[3]{\ensuremath{{#1}\ {#3}\ {#2}}}
\def\eEquiv#1#2{\isEquiv{#1}{#2}{\ePDL}}
\def\eklasse#1{\Klasse{#1}{\ePDL}}
\def\efaktor#1{\Faktor{#1}{\ePDL}}
\def\IK#1#2{\ensuremath{\mathbf{I}_{\K}(#1\,|\,#2)}}
\def\IKa#1#2{\ensuremath{\mathbf{I}_{\K}(#1,#2)}}
\def\IM#1#2{\ensuremath{\mathbf{J}_{\Mo}(#1\,|\,#2)}}
\def\IMa#1#2{\ensuremath{\mathbf{J}_{\Mo}(#1,#2)}}

\usepackage{fancyhdr}
\begin{document}
\maketitle
\begin{abstract}
The interpretation of propositional dynamic logic (PDL) through Kripke models requires  the relations constituting the interpreting Kripke model to closely observe  the syntax of the modal operators. This poses a significant challenge for an interpretation of PDL through stochastic Kripke models, because the programs' operations do not always have a natural counterpart in the set of stochastic relations. We use rewrite rules for  building up an interpretation of PDL. It is shown that each program corresponds to an essentially unique irreducible tree, which in turn is assigned a predicate lifting, serving as the program's interpretation. The paper establishes and studies this interpretation. It discusses the expressivity of  probabilistic models for PDL and relates properties like logical and behavioral equivalence or bisimilarity to the corresponding properties of a Kripke model for a closely related non-dynamic logic of the Hennessy-Milner type. 
\end{abstract}
\section{Introduction}
\label{sec:intro}

The interpretation of propositional dynamic logic (PDL) through Kripke models requires, as is customary in modal logics,  the relations in the interpreting Kripke model to closely observe  the syntactic properties of the modal operators~\cite[Section 2.4]{Blackburn-Rijke-Venema}. For example, the nondeterministic choice $\pi\cup\pi'$ of programs $\pi$ and $\pi'$ is usually interpreted through relation $R_{\pi\cup\pi'}$ which satisfies $R_{\pi\cup\pi'} = R_{\pi}\cup R_{\pi'}$, and the relation for the indefinite iteration $\pi^*$ should satisfy $R_{\pi^*} = R_{\pi}^*$.

This poses a significant challenge for an interpretation of PDL through stochastic Kripke models, because the programs' operations do not always have a natural counterpart in the set of stochastic relations. Clearly, operations like $K_{\pi}\cup K_{\pi'}$ or $K^*_{\pi}$ hardly make sense for transition probabilities $K_{\pi}$ and $K_{\pi'}$. In addition, an interpretation of PDL observes usually some tacit assumptions on the ``static'' semantics like $\pi_1;(\pi_2\cup\pi_3) = \pi_1;\pi_2 \cup \pi_1;\pi_3.$ 

We convert these implicit assumptions into rewrite rules. This permits building up an interpretation of PDL through terms in an algebra. Because we have to cater for the indefinite iteration of a program, the algebra admits an operator of infinite arity. It is shown that each program corresponds to an essentially unique irreducible tree, which in turn is assigned a natural transformation, serving as the programm's interpretation. Some technical problems have to be overcome due to the observation that the interpretation of the indefinite iteration ---~the counterpart of the \texttt{while}-loop~--- requires a base space which is closed under the well-known Souslin operation from set theory. This is in particular inconvenient when the state space is assumed to be Polish: these spaces are closed under this operation only if they are finite. Hence previous results on the stochastic coalgebraic interpretation of modal logics are difficult to apply.

The paper discusses the expressivity of these models and relates properties like logical and behavioral equivalence or bisimilarity to the corresponding properties of a Kripke model for a closely related non-dynamic logic of the Hennessy-Milner type. 

We will  in Section~\ref{sec:progs} have a look at term rewriting for programs, producing an irreducible tree from a program. This tree is well-founded, hence has no infinitely long paths, but it may have nodes with an infinite fan-out; these are exactly the nodes which correspond to the \texttt{while}-loop. We are able to produce an interpretation from an irreducible tree, provided we can interpret primitive programs, and we know how to handle the choice and the iteration operator. These operators are given through natural transformations for the Borel functor. We study these transformations in Section~\ref{sec:trans} together with some properties of the underlying measurable spaces; this is becomes necessary because the presence of the iteration operator complicates the measurable structure of the validity sets, as shown in~\cite{EED-PDL-TR}. Sections~\ref{sec:interpr} and~\ref{sec:theLogic} deal with models and interpretations: we first define the usual Kripke models and extend them to incorporate natural transformations. They will then help to define the semantics of PDL formulas. On the other hand, a simple modal logic of the Hennessy-Milner type is defined,  the modal operators being given through the primitive programs. These logics are compared and help to give some insight into the question of expressivity; again, we have to be a bit careful because the case \emph{Bisimilarity Vs. Behavioral Equivalence} makes some topological assumptions mandatory for a successful discussion. This requires extending the notion of a model in Section~\ref{sec:g-models}  for capturing fully the development discussed to far. A satisfactory answer on the equivalence of all three variants of expressivity can be given under the assumption that the respective sets of atomic expressions and  of primitive programs both are countable. Finally, Section~\ref{sec:conclusion} wraps it all up and suggests further work.  
\section{Programs}
\label{sec:progs}

The modalities for PDL are given through a simple grammar which is intended to model programs. When interpreting the logic through a Kripke model, the problem arises that not each modal operator has a relation associated with it. Associating a relation with each primitive program and working in a monad permits interpreting the composition of primitive programs through Kleisli composition, but there is no provision for interpreting operators like the nondeterministic choice or the indefinite iteration. These interpretations have to be constructed explicitly. In order to be able to do this, we study the set of all programs first, introducing rewrite rules and equations for reducing programs to a simpler, more manageable form.

\medskip

The grammar for programs over the set $\Ur$ of primitive programs is given by 
\begin{equation*}
\pi ::= \varrho \mid \pi_1\cup\pi_2 \mid \pi_1;\pi_2 \mid \pi^*
\end{equation*}
with $\varrho\in\Ur$. We assume that the empty program $\epsilon$ is a member of $\Ur$. The set $\Pro$ of programs over $\Ur$ is perceived as the term algebra over the constants $\Ur$ with the unary operation $\cdot^*$ and the binary operations $\{;, \cup\}$. Program $\pi_1\cup\pi_2$ is the nondeterministic choice of programs $\pi_1$ and $\pi_2$, $\pi_1;\pi_2$ is sequential composition, and $\pi^*$ is indefinite iteration: executing $\pi^*$ entails executing $\pi$ $k$ times with $k \geq 0$. 

We assume that we have an operation $\infOp$ of infinite arity. Denote the term algebra for the operators $\{;, \cup, *, \infOp\}$ over $\Ur$ by $\ProExt$. The free semigroup over $\Ur$ with respect to sequential program composition (the \emph{basic blocks} of compiler construction) is denoted by $\wrd{\Ur}$. 

Each program $\pi$ is given an ordinal number $\ord(\pi)$ as its weight. It is defined recursively through
\begin{equation*}
\ord(\pi) := 
\begin{cases}
1,& \text{if } \pi=\epsilon,\\
2,& \text{if } \pi\in\Ur\setminus\{\epsilon\},\\
\ord(\pi_1)\cdot\ord(\pi_2),& \text{if } \pi = \pi_1;\pi_2,\\
\ord(\pi_1) + \ord(\pi_2) + 1,& \text{if } \pi = \pi_1\cup\pi_2,\\
\sup_{k\in\Nat} \ord(\pi_1^k), & \text{if } \pi = \pi_1^*.
\end{cases}
\end{equation*}
Here $\pi^k$ is defined as the $k$-fold iteration of $\pi$, thus
\begin{equation*}
\pi^k := 
\begin{cases}
\epsilon& \text{if } k = 0,\\
\pi^{k-1};\pi& \text{otherwise}.
\end{cases}
\end{equation*}
Form the definition it is clear that $\ord(\pi) < \infty$ iff $\pi$ does not contain any iteration, i.e., a subexpression of the form $\pi_1^*$. 

The static semantics of program composition is usually given through informal rules: executing $\pi_1;(\pi_2\cup\pi_3)$, i.e., executing first $\pi_1$ and then choosing between $\pi_2$ and $\pi_3$ should be the same as choosing between $\pi_1;\pi_2$ and $\pi_1;\pi_3$, or executing $\pi_1;\pi_2^*; \pi_3$ should give the choice of executing $\pi_1;\pi_3$ (i.e., not executing $\pi_2$ at all), and $\pi_1;\pi_2;\pi_2^*;\pi_3$ (i.e., executing $\pi_2$ at least once in the context of $\pi_1$ and $\pi_3$). It helps for a coalgebraic interpretation to have a formal specification of these rules. We propose to use rewrite rules for this, augmented by equations which state properties like associativity). 

We introduce these rewrite rules (in order to avoid parentheses, we assume that operator $;$ binds tighter than the operator $\cup$):
\begin{alignat*}{2}
\DistrL\quad& x;(y\cup z)\quad&\to\quad&x;y\cup x;z\\
\DistrR\quad&(x\cup y); z\quad&\to\quad&x;z \cup y;z\\
\Emb\quad&x^*\quad&\to\quad&\epsilon;x^*;\epsilon  \\
\Star\quad&x;y^*;z\quad&\to\quad&x;y\cup x;y;y^*;z  \\
\end{alignat*}
These are the equations:
\begin{xalignat*}{2}
\IdEpsl\quad&\epsilon;x\quad&\approx\quad &x\\
\IdEpsr\quad&x;\epsilon\quad&\approx\quad &x\\
\AssS\quad& x;(y ;z)\quad&\approx\quad&(x; y);z\\
\AssU\quad&x\cup (y \cup z)\quad&\approx\quad&(x\cup y) \cup z\\
\Comm\quad&x \cup y\quad&\approx\quad&y \cup x\\
\Idemp\quad&x \cup x\quad&\approx\quad&x\\
\InfDisj\quad&\infOp\langle x_k| k \geq 0\rangle\quad&\approx\quad&x_0 \cup \infOp\langle x_{k+1}| k \geq 0\rangle\\
\Transp\quad&\infOp\bigl\langle\infOp\langle x_{k, \ell}| k \geq 0\rangle|\ell\geq 0\bigr\rangle\quad&
\approx\quad&
\infOp\bigl\langle\infOp\langle x_{k, \ell}| \ell\geq 0\rangle|k \geq 0\bigr\rangle
\end{xalignat*}

The first group of equations states that $\epsilon$ plays the role of the program \texttt{skip}, and that choice as well as sequential composition are associative; choice is commutative as well. The last group deal with the operator $\infOp$ which is assumed to be the implementation of the indefinite iteration. 
Equation $\InfDisj$ is akin to an infinite associative law: considering an infinite choice of programs is the same as considering the choice between the first one and the rest. Equation $\Transp$ says that $\pi_1^*;\pi_2^*$ can be interpreted as either $\pi_1$ terminating after a finite number of steps followed by $\pi_2^*$ or as $\pi_1^*$ followed by a finite number of executions of $\pi_2$. 
 
The set $X$ of variables is assumed to be a countable set. As usual, a substitution $\sigma$ is a map from $X$ to $\Pro$ which is extended accordingly. 

Following~\cite{Dersh:Rewrite}, a term $\alpha\in\ProExt$ is perceived as an ordered tree, each node in which has address $\adr{a}$ in the Dewey notation (the node with address $\adr{a} = \adr{0.1.3}$ is reached through taking the leftmost son of the root, then its second son and finally the fourth offspring); the subtree of $\alpha$ rooted at the node which has the address $\adr{a}$ is denoted by \subTree{\alpha}{a}. Denote by $\subst{\alpha}{\gamma}{a}$ denotes the tree in which the subtree of $\alpha$ which is rooted at $\adr{a}$ is replaced by the tree associated with term $\gamma$. 

We say that $\alpha\To\beta$ iff there exists a rule $l\to r$, a position $\adr{a}$ and a substitution $\sigma$ such that $\subTree{\alpha}{a} = \sigma(l)$ and $\subst{\alpha}{\sigma(r)}{a} = \beta$.  The reflexive-transitive closure of $\To$ is denoted as usual by $\To^*$. Call $\alpha\in\ProExt$ \emph{irreducible} iff there is no $\beta\in\ProExt$ with $\alpha\To^*\beta$ and $\beta \not=\alpha$. 

Denote by $\equiv$ the congruence defined by $\approx \cup \To$ on $\ProExt$, thus $\equiv$ is the smallest equivalence relation on $\ProExt$ which is compatible with the operations $\{;, \cup, *, \infOp\}$ on $\ProExt$ and which contains the relation  $\approx \cup \To$. The canonical projection which assigns $\alpha\in\ProExt$ its class $\klasse{\alpha}$ is denoted by $\fMap{\equiv}: \ProExt\to\Faktor{\ProExt}{\equiv}$. 

The following statement shows that rewriting a program with finite weight always terminates. It does not give, however, a unique result, the result is rather determined uniquely up to $\equiv$ (which is not surprising given, e.g., associativity, commutativity and idempotence of the nondeterministic choice). 

\BeginLemma{red-starfree}
Let $\pi\in\Pro$ be a program with $\ord(\pi) < \infty$. Then there exists $F\subseteq \wrd{\Ur}$ finite with 
$
\pi \equiv \bigcup F.
$
If
$
\pi \equiv \bigcup F'
$
for some  finite $F'\subseteq \wrd{\Ur}$, then 
$
\Bild{\fMap{\equiv}}{F} = \Bild{\fMap{\equiv}}{F'}.
$
\EndLemma

\BeginProof
Note that 
$
\ord\bigl(\pi_1;(\pi_2\cup\pi_3)\bigr) > \ord(\pi_1;\pi_2\cup\pi_1;\pi_3),
$
(see~\cite[p.~270]{Dersh:Rewrite}), similarly for rule $\DistrR$. Because $\ord(\pi) < \infty$, any application of the rewrite rules $\DistrL$ and $\DistrR$ terminates. Thus
$
\pi \equiv \bigcup F
$
for some $F\subseteq \wrd{\Ur}$ finite. Uniqueness up to $\equiv$ is established by induction on the structure of $\pi$.  
\EndProof

\medskip

These are some properties of irreducible elements of $\ProExt$.

\BeginLemma{prop-irred}
Denote by $\ProIrr$ the set of irreducible elements in $\ProExt$.

\begin{enumerate}[a)]
\item\label{prop-irred-a} $\ProIrr$ is closed under the operators $\cup$ and $\infOp$.
\item\label{prop-irred-b} If $\beta_1, \beta_2\in\ProIrr$, there exists $\beta'\in\ProIrr$ such that 
$
\beta_1;\beta_2\equiv\beta'.
$
\item\label{prop-irred-c} If $\pi\in\Pro$ with $\ord(\pi) < \infty$, then $\pi$  is irreducible iff there exists $F\subseteq \wrd{\Ur}$ with 
$
\pi \stackrel{\AssS}{=} \bigcup F,
$ 
$\stackrel{\AssS}{=}$ denoting equality modulo associativity of operator $;$. 
\end{enumerate}
\EndLemma

\BeginProof
1.
It is clear that $\ProIrr$ is closed under $\cup$ because there is no rewrite rule which has $\cup$ as its main operator on its left hand side. It is also clear that $\ProIrr$ is closed under the infinite operator $\infOp$, because each transformation of such a term is pushed into its components. Each element of $\wrd{\Ur}$ is irreducible, so is their finite union. From this follows the claim for programs of finite rank.

2.
Note that the syntax tree associated with an element of $\ProExt$ is well formed, since it does not have paths of infinite length. An easy induction on the tree for $\beta\in\ProIrr$ shows that if $\varrho\in\wrd{\Ur}$, then there exists $\beta'\in\ProIrr$ with $\varrho;\beta \equiv \beta'$. 

In fact, if $\beta = \pi\in\Pro$ with $\ord(\pi) < \infty$, or if $\beta \equiv \beta_1\cup\beta_2$ with irreducible $\beta_1, \beta_2$, the claim follows easily. If we can write $\beta  \equiv \beta_1;\beta_2$ then irreducibility of $\beta$ implies irreducibility of $\varrho;\beta$. Finally, assume that 
$
\beta \equiv \infOp \langle \beta_k| k \geq 0\rangle,
$
then all $\beta_k$ are irreducible, and 
$
\varrho;\beta \equiv \infOp \langle \varrho;\beta_k| k \geq 0\rangle.
$ 
For $\varrho; \beta_k$ we find $\beta'_k$ with $\varrho; \beta_k\equiv\beta'_k$ by induction hypothesis, so that 
$
\beta \equiv \beta' := \infOp \langle \beta'_k| k \geq 0\rangle
$
with $\beta'\in\ProIrr$. 

3.
We show now that $\beta_1, \beta_2\in\ProIrr$ implies the existence of $\beta'\in\ProIrr$ with $\beta_1;\beta_2\equiv\beta'$ by induction on the syntax tree for $\beta_1$. If this tree is finite, then parts 1. and 2. show that $\beta_1;\beta_2 \equiv\bigcup_{\varrho\in F}\varrho;\beta_2 \equiv\bigcup_{\varrho\in F}\beta_\varrho$ with $\beta_\varrho\in\ProIrr$ for some finite $F\subseteq\wrd{\Ur}$. 
Assume $\beta_1 = \infOp\langle \beta_{1, k}| k \geq 0\rangle$. By the induction hypothesis we know that for each $k$ there exists $\beta_k'\in\ProIrr$ such that $\beta_{1, k};\beta_2 \equiv \beta_k'$, so that 
$
\beta_1;\beta_2 \equiv \infOp\langle \beta_k'| k \geq 0\rangle,
$
the latter being irreducible. 

If the tree for $\beta_1$ is infinite and has the operator $;$ as its root, say
$
\beta_1 = \beta_{1, a};\beta_{1, b},
$
then at least one of the trees for $\beta_{1, a}$ or $\beta_{1, b}$ is infinite. Assume without loss of generality that 
$
\beta_{1, a} = \infOp\langle \beta_{1, a, k}| k \geq 0\rangle,
$
then
$ 
\beta_1 \equiv \infOp\langle\beta_{1, a, k};\beta_{1, b}| k \geq 0\rangle.
$
Consequently, the induction hypothesis  may be applied through the same argumentation as above. 
\EndProof

This has as an immediate consequence that each program is equivalent to an irreducible one (which may have infinite branches).

\BeginCorollary{each-irred}
Given a program $\pi\in\Pro$, there exists $\beta\in\ProIrr$ such that $\pi\equiv\beta$.
\EndCorollary

\BeginProof
The proof proceeds by induction on $\ord(\pi)$. If $\ord(\pi) < \infty$, the assertion follows from Lemma~\ref{prop-irred}, part~\ref{prop-irred-c}. Now let $\pi$ with $\ord(\pi) = \infty$ be given, and assume that the assertion is established for all programs $\pi'$ with $\ord(\pi') < \ord(\pi)$. If $\pi = \pi_1\cup\pi_2$ or $\pi = \pi_1^*$, the assertion follows from the induction hypothesis together with part~\ref{prop-irred-a} in Lemma~\ref{prop-irred}. If, however, $\pi = \pi_1;\pi_2$, we apply the induction hypothesis to $\pi_1$ and $\pi_2$, the assertion then follows from part~\ref{prop-irred-b} in Lemma~\ref{prop-irred}.
\EndProof

\medskip

Because $\equiv$ is a congruence, these operations on $\faktor{\ProExt}$ are well defined:
\begin{align*}
\klasse{\pi_1}\sqcup\klasse{\pi_2} & := \klasse{\pi_1\cup\pi_2},\\
\bigsqcup \bigl\langle\klasse{\pi_k}| k \geq 0\bigr\rangle & := \klasse{\infOp\langle \pi_k| k \geq 0\rangle}
\end{align*}

Define the map $\Theta: \Pro \to \Faktor{\ProExt}{\equiv}$ inductively on the weight of program $\pi$ as follows.
\begin{enumerate}[a)]
\item\label{pi-finite} If $\ord(\pi) < \infty$, put 
\begin{equation*}
\Theta(\pi) := \bigsqcup\{\klasse{\varrho} \mid \varrho \in F\}
\end{equation*} 
with 
$
\pi \equiv \bigcup F
$
and $F\subseteq \wrd{\Ur}$ according to Lemma~\ref{red-starfree}.
\item\label{pi-union} Proceeding inductively, assume that $\Theta(\pi_1)$ and $\Theta(\pi_2)$ are defined, then put 
\begin{equation*}
\Theta(\pi_1\cup\pi_2) := \Theta(\pi_1)\sqcup\Theta(\pi_2).
\end{equation*}
\item\label{pi-semi} Continuing with an inductive definition, assume that $\pi = \pi_1;\pi_2$ with $\ord(\pi)$ not finite. We distinguish there cases
\begin{enumerate}[(i)]
\item\label{pi-semi-1} $\ord(\pi_1)$ is finite. Since $\ord(\pi_1;\pi_2)$ is not finite, we can represent $\ord(\pi_2)$ through $\mathfrak{m}_0 + k$, where $\mathfrak{m}_0$ is a limit ordinal and $k$ is finite. Thus 
$
\pi_2 \equiv \pi_{2, a}\cup\pi_{2, b}
$ 
with $\ord(\pi_{2, a}) = \mathfrak{m}_0$ and $\ord(\pi_{2, b}) = k$. Then  
$
\pi_{2, a} \equiv \hat{\pi};\hat{\pi}_{2, a}
$
with $\ord(\hat{\pi})$ finite and $\hat{\pi}_{2, a} = \pi_{2, c}^*$. This is so since $\ell\cdot \mathfrak{m} = \mathfrak{m}$ for any finite $\ell$ and any limit ordinal $\mathfrak{m}$. Thus
\begin{align*}
\pi 
&\equiv 
\pi_1;(\hat{\pi};\pi_{2, c}^* \cup \pi_{2, b})\\
&\equiv (\pi_1;\hat{\pi});\pi_{2, c}^* \cup \pi_1;\pi_{2, b}.
\end{align*}
Because both $\ord(\pi_1;\hat{\pi})$ and $\ord(\pi_1;\pi_{2, b})$ are finite, and since $\ord(\pi_{2, c}^k) < \ord(\pi_{2, c}^*)$, $\Theta$ is defined for these arguments, and we put
\begin{equation*}
\Theta(\pi) := \bigsqcup\langle\Theta(\pi_1;\hat{\pi};\pi_{2, c}^k)| k \geq 0\rangle\sqcup\Theta(\pi_1;\pi_{2, b}).
\end{equation*}
\item\label{pi-semi-2} $\ord(\pi_2)$ is finite. We find $F \subseteq \wrd{\Ur}$ finite with 
$
\pi \equiv \bigcup\{\pi_1;\varrho\mid \varrho \in F\}.
$
Similar to the case above we represent $\pi_1 \equiv \pi_0;\pi_{1, a}^*\cup\pi_{1, b}$ with both $\ord(\pi_0)$ and $\ord(\pi_{1, b})$ finite. Hence 
$
\pi_0 \equiv \bigcup\{\varrho' \mid \varrho' \in G\}
$
for some finite $G \subseteq \wrd{\Ur}$. Then define
\begin{equation*}
\Theta(\pi) := \bigsqcup_{\varrho\in F}\bigsqcup_{\varrho'\in G}\langle\Theta(\varrho';\pi_{1, a}^k;\varrho)| k \geq 0\rangle
\sqcup \Theta(\pi_{1, b};\pi_2).
\end{equation*}
\item\label{pi-semi-3} Both $\ord(\pi_1)$ and $\ord(\pi_2)$ are not finite. Represent
\begin{align*}
\pi_1 & \equiv \pi_{1, a}; \pi_{1, b}^* \cup \pi_{1, c},\\
\pi_2 & \equiv \pi_{2, a}; \pi_{2, b}^* \cup \pi_{2, c}
\end{align*}
with $\ord(\pi_{1, a}), \ord(\pi_{1, c}),\ord(\pi_{2, a}),\ord(\pi_{2, c})$ finite. Apply the rules $\DistrL$ and $\DistrR$ to obtain
\begin{equation*}
\pi_1;\pi_2 \equiv  \pi_{1, a};\pi_{1, b}^*;\pi_{2, a};\pi_{2, b}^*\cup \pi_{1, c};\pi_{2, a};\pi_{2, b}^*\cup \pi_{1, a};\pi_{1, b}^*;\pi_{2, c}\cup \pi_{1, c};\pi_{2, c}.
\end{equation*} 
Because we may represent 
$
\pi_{1, a} = \bigcup \{\varrho \mid \varrho \in F\} 
$
and
$
\pi_{2, a} = \bigcup \{\varrho' \mid \varrho' \in F'\}
$
for some finite $F, F'\subseteq \wrd{\Ur}$, we may and do assume that $\pi_{1, a}, \pi_{2, a}\in\wrd{\Ur}$. Put
\begin{align*}
\Theta(\pi_{1, a};\pi_{1, b}^*;\pi_{2, a};\pi_{2, b}^*)
& := 
\bigsqcup_{k\geq 0}\bigsqcup_{\ell\geq 0}\Theta(\pi_{1, a};\pi_{1, b}^k;\pi_{2, a};\pi_{2, b}^\ell)\\
\bigl(& = \bigsqcup_{\ell\geq 0}\bigsqcup_{k\geq 0}\Theta(\pi_{1, a};\pi_{1, b}^k;\pi_{2, a};\pi_{2, b}^\ell)\bigr)
\end{align*}
Because
$
\max\{\ord(\pi_{1, c};\pi_{2, a};\pi_{2, b}^*), \ord(\pi_{1, a};\pi_{1, b}^*;\pi_{2, c}), \ord(\pi_{1, c};\pi_{2, c})\} < \ord(\pi), 
$
we may now define
\begin{equation*}
\Theta(\pi) := \Theta(\pi_{1, a};\pi_{1, b}^*;\pi_{2, a};\pi_{2, b}^*)\sqcup \Theta(\pi_{1, c};\pi_{2, a};\pi_{2, b}^*)\sqcup \Theta(\pi_{1, a};\pi_{1, b}^*;\pi_{2, c})\sqcup \Theta(\pi_{1, c};\pi_{2, c}).
\end{equation*}
\end{enumerate}
\end{enumerate}

The construction shows that $\pi\equiv\beta$ for $\beta\in\ProIrr$ entails $\beta\in\Theta(\pi)$, thus we obtain from Corollary~\ref{each-irred}

\BeginProposition{well-defined}
$\Theta: \Pro \to \Faktor{\ProExt}{\equiv}$ is well defined. 
\QED
\EndProposition

Summarizing, we construct for a program $\pi\in\Pro$ an equivalence class which contains an irreducible element of $\ProExt$. Such an irreducible program is composed of the choice operator and the explicit form of the indefinite iteration. The primitive programs appear only in the form of basic blocks $\varrho_1;\dots;\varrho_k$ with $\varrho_1, \dots, \varrho_k\in\Ur$. 

Consequently, an interpretation of a logic carrying programs for modalities will have to cater for the respective interpretation of the choice operator, the explicit form of the indefinite iteration, and the basic blocks. The latter ones can be composed from the interpretation of the primitive programs for example in those cases that are given by a monad, where composition of programs may be modelled through Kleisli composition~\cite{Moggi-Inf+Control}. 

Instead of providing after the preparations above a general coalgebraic interpretation through a monad over the category of sets now, we propose an interpretation through stochastic relations (which offers its own idiosyncrasies in turn). 
\section{Transformations}
\label{sec:trans}

We collect for the reader's convenience some techniques and tools from set theory and probability, in particular techniques for working with $\sigma$-algebras and their completion.

\subsection{Measurability}
\label{sec:measur}

A \emph{measurable space} $S$ is a set, again denoted by $S$, together with a Boolean $\sigma$-algebra $\Borel{S}$, thus $\Borel{S}$ is an algebra of sets which is also closed under countable unions. Denote for a set $\mathcal{A}$ of subsets of a set $S$ by $\sigma(\mathcal{A})$ the smallest $\sigma$-algebra containing $\mathcal{A}$. 

A map $f: S\to T$ is called \emph{$\Borel{S}$-$\Borel{T}$-measurable} (or just \emph{measurable}, if the context is clear) iff the inverse image of each Borel set in $T$ is a Borel set in $S$, or, formally, iff 
\begin{equation*}
\InvBild{f}{\Borel{T}} := \{\InvBild{f}{C} \mid C \in \Borel{T}\} \subseteq \Borel{S}.
\end{equation*}
If $\Borel{T} = \sigma(\mathcal{A})$, then $f: S \to T$ is measurable iff $\InvBild{f}{A}\in \Borel{S}$ for all $A\in \mathcal{A}$. 

The real numbers always carry the Borel sets $\Borel{\Real}$ as a $\sigma$-algebra, where 
\begin{equation*}
\Borel{\Real} := \sigma(\{G \subseteq \Real \mid G \text{ open}\}) = \sigma(\{]a, b[ \mid a, b \in \Real, a < b\}).
\end{equation*} 
Let $\SubProb{S}$ be the set of all subprobabilities on measurable space $S$, then $\Borel{\SubProb{S}}$ will be the weak-*-$\sigma$-algebra, i.e., the smallest $\sigma$-algebra on $\SubProb{S}$ which makes all the evaluations 
$
ev_A: \mu \mapsto \mu(A)
$
Borel-measurable. Then 
\begin{equation*}
\Borel{\SubProb{S}} = \sigma(\{\basB_{q, A} \mid q \in \Rational, A \in \Borel{S}\})
\end{equation*}
with 
\begin{equation*}
\basB_{q, A} := ev_A^{-1}\bigl[]-\infty, q[\bigr] = \{\mu\in\SubProb{S} \mid \mu(A) < q\}.
\end{equation*}
A \emph{stochastic relation} $K: S \Trans T$ between the measurable spaces $S$ and $T$ is a Borel measurable map from $S$ to $\SubProb{T}$; sometimes stochastic relations are called \emph{transition subprobabilities}. Thus $K: S \Trans T$ is a stochastic relation iff $K(s)$ is a subprobability on the measurable space $T$ for each $s\in S$ such that $s \mapsto K(s)(B)$ is a $\Borel{S}$-measurable function for each $B\in\Borel{T}$.  

Denote by $\M$ the category of measurable spaces with measurable maps as morphisms, and by $\N$ the category of all $\sigma$-algebras with maps. The \emph{Borel functor $\B: \M\to\N$} assigns to each measurable space its Borel sets, and to a morphism $f: S\to T$ its inverse image $f^{-1}: \Borel{T}\to\Borel{S}$. Thus $\B$ is a contravariant functor. This has been discussed extensively in~\cite{EED-CS-Survey,EED-CoalgLogic-Book}. Given a morphism $f: S \to T$ in category $\M$, we obtain a morphism $\SubProb{f}: \SubProb{S}\to\SubProb{T}$ in $\M$ upon defining 
\begin{equation*}
\SubProb{f}(\mu)(B) := \mu(\InvBild{f}{B})
\end{equation*}
for $\mu\in\SubProb{S}$ and $B\in\Borel{T}$. $\SubProb{f}$ is $\Borel{\SubProb{S}}$-$\Borel{\SubProb{T}}$-measurable because
\begin{equation*}
\InvBild{\SubProb{f}}{\basB_{q, B}} = \basB_{q, \InvBild{f}{B}}
\end{equation*}
holds for each real $q$ and each measurable set $B\in\Borel{T}$. Functor $\SubProbSenza$ is the functorial part of a monad which is sometimes called the \emph{Giry monad}~\cite{Giry,EED-KleisliMorph,EED-Book}. 

Let $K: S \Trans S$ and $L: T \Trans T$ be stochastic relations for the measurable spaces $S$ and $T$, then a measurable map $f: S \to T$ is called a \emph{morphism} $K\to L$ iff 
$
L \circ f = \SubProb{f}\circ K
$
holds, rendering the diagram
\begin{equation*}
\xymatrix{
S\ar[rr]^f\ar[d]_K && T\ar[d]^L\\
\SubProb{S}\ar[rr]_{\SubProb{f}} && \SubProb{T}
}
\end{equation*} 
commutative. Expanded, this means that
\begin{equation*}
L(f(s))(B) = K(s)(\InvBild{f}{B})
\end{equation*}
holds for each state $s\in S$ and each measurable set $B\in\Borel{T}.$

We will need this technical statement for transformations when considering runs of primitive programs below.
\BeginLemma{CVT}
Let $S$ and $T$ be measurable spaces, $f: S \to T$ be a measurable map. Assume that $g: T \to \Real$ is measurable and bounded.
\begin{enumerate}[a.]
\item\label{CVT-a} For any $\mu\in\SubProb{S}$ 
\begin{equation*}
\int_T g(y)\ \SubProb{f}(\mu)(dy) = \int_S (g \circ f)(x)\ \mu(dx).
\end{equation*}
\item\label{CVT-b}
If $f: K \to L$ is a morphism for the stochastic relations $K: S \Trans S$ and $L: T \Trans T$, then
\begin{equation*}
\int_T g(y)\ L(f(s))(dy) = \int_S (g \circ f)(x)\ K(s)(dx).
\end{equation*}
\end{enumerate}
\EndLemma

\BeginProof
The formula in part~$\ref{CVT-a}.$ is the classical Change of Variables Formula, see~\cite[Lemma 1.6.20]{EED-CoalgLogic-Book}. 
Part~$\ref{CVT-b}.$ is an immediate consequence: because 
$
L(f(s)) = \SubProb{f}(K(s)),
$
we may write
\begin{equation*}
\int_T g(y)\ L(f(s))(dy)
=
\int_T g(y)\ \bigl(\SubProb{f}(K(s))\bigr)(dy)
=
\int_S g(f(x))\ K(s)(dx),
\end{equation*}
the last equation being due to part~$\ref{CVT-a}.$
\EndProof

\subsection{The Souslin Operation}
\label{sec:souslin}

When interpreting the indefinite iteration $\pi^*$ of  program $\pi$, we will be faced with the problem that validity sets for formulas formed using $\pi^*$ will be using uncountable unions. Thus these validity sets may not be measurable, because measurability always assumes countable operations. There is, however, a broad class of measurable spaces which permit uncountable operations in restricted form; by a completion operation, each measurable space can be embedded into such a space. This restricted form is described by the Souslin operation, which will be introduced now.

A measurable space $S$ is closed under the \emph{Souslin operation} iff, whenever 
$
\{A_v \mid v \in\wrd{\Nat_0}\} \subseteq \Borel{S}
$
is a family of measurable sets indexed by finite sequences of natural numbers, we have
\begin{equation*}
\bigcup_{\alpha\in\Nat_0^{\Nat}}\bigcap_{n\in\Nat} A_{\alpha|n} \in \Borel{S},
\end{equation*}
where $\alpha|n$ are the first $n$ elements of sequence $\alpha$. This is sometimes called \emph{operation $\mathcal{A}$} on the \emph{Souslin scheme $\{A_v \mid v \in\wrd{\Nat_0}\}$}~\cite[XI.5]{Kuratowski-Mostowski}. 

Define for the measurable space $S$ and a subprobability $\mu\in\SubProb{S}$ its \emph{$\mu$-completion $\cpl{S}{\mu}$} through 
\begin{equation*}
A \in \Borel{\cpl{S}{\mu}} \Leftrightarrow \exists A_0, A_1\in\Borel{S}: A_0\subseteq A \subseteq A_1\text{ and } \mu(A_1\setminus A_0) = 0.
\end{equation*}slin
Thus all sets which differ from a Borel set by a set on $\mu$-measure $0$ are added to the Borel sets; the underlying set remains unchanged. Then $\Borel{\cpl{S}{\mu}}$ is a $\sigma$-algebra again. If $M\subseteq\SubProb{S}$ is a non-empty set of subprobabilities on $S$, put
\begin{equation*}
\Borel{\cpl{S}{M}} := \bigcap_{\mu\in M} \Borel{\cpl{S}{\mu}}.
\end{equation*}

\BeginDefinition{completion}
$\cpl{S}{M}$ is called the \emph{$M$-completion of $S$}, $\cpl{S}{\SubProb{S}}$ is called the \emph{universal completion of $S$} and is denoted by $\cpl{S}{}$. 
\EndDefinition

The important property reads
\BeginProposition{compl-sou}
The measurable space $\cpl{S}{M}$ is closed under the Souslin operation for every $\emptyset\not=M\subseteq\SubProb{S}$. 
\EndProposition

\BeginProof
\cite[Theorem 3.5.22]{Srivastava}.
\EndProof

Measurability of maps carries over to the completion.

\BeginLemma{remains-meas}
Given measurable spaces $S$ and $T$, and assume that $f: S \to T$ is $\Borel{S}$-$\Borel{T}$-measurable.
\begin{enumerate}[a.]
\item Let $M \subseteq \SubProb{S}, N \subseteq \SubProb{T}$ such that $\SubProb{f}(\mu) \in N$ for all $\mu\in M$. Then $f$ is $\Borel{\cpl{S}{M}}$-$\Borel{\cpl{T}{N}}$-measurable.
\item $f$ is $\Borel{\cpl{S}{}}$-$\Borel{\cpl{T}{}}$-measurable. 
\end{enumerate}  
\EndLemma

\BeginProof
\cite[Proposition 4.3]{EED-PDL-TR}.
\EndProof

We note for later use that a stochastic relation can be extended to the completion of a measurable space as well, provided the measurable space is \emph{separable}. This means that the Borel sets are countably generated, formally: 

\BeginDefinition{separable}
$S$ is called \emph{separable} iff there exists a countable family $\mathcal{A}_0$ of subsets of $S$ such that $\Borel{S} = \sigma(\mathcal{A}_0)$. 
\EndDefinition

For example, $\Real$ is separable, so is every measurable space that has as Borel sets the $\sigma$-algebra generated by the open sets of a topological space with a countable base. Polish spaces are important special case: call a second countable topological space \emph{Polish} iff the topology can be metrized with a complete metric. The Borel sets of a Polish space are countably generated, so that a measurable space generated from a Polish space is separable; the natural topology on the reals is Polish. A measurable space generated from a Polish space is called a \emph{Standard Borel} space (hence discussing a Standard Borel space, we are not interested in its topological but rather in its measurable structure). 

The following proposition shows why separable measurable spaces are of interest to us. We will use it later for completing models (but maintaining expressivity).

\BeginProposition{extension-to-compl}
Let $S$ be a separable measurable space, $K: S \Trans S$ be a stochastic relation on $S$. Then there exists a unique stochastic relation $\overline{K}: \Borel{\cpl{S}{}}\Trans \Borel{\cpl{S}{}}$ extending $K$. Let $L$ be another stochastic relation defined over a separable measurable space. If $f: K \to L$ is a morphism, then $f: \overline{K}\to\overline{L}$ is a morphism.
\EndProposition

\BeginProof
\cite[Proposition 7.10, Corollary 7.6]{EED-PDL-TR}
\EndProof

\subsection{Natural Transformation}
\label{sec:nat-trans}

The category of all measurable spaces which are closed under the Souslin operation is denoted by $\V$, the restriction of functor $\B$ to $\V$ is again denoted by $\B$.

Denote by $\S$ the category of stochastic relations; it has pairs $\langle S, R\rangle$ as objects and the morphisms defined above as morphisms. Define functor $\BS$ on $\S$ through functor $\B$ by defining $\BS := \B\circ\U$ with $\U: \S\to\M$ as the forgetful functor; hence $\BS(S, R) = \B(S)$, and $\BS$ acts on morphisms accordingly. ``Daggering'' a functor will compose it with the forgetful functor $\U$. 

The constant functor assigning each measurable space the rationals between $0$ and $1$ 
is also denoted by $\Q$. Let $\NQ$ be the category which has all maps $\Rational\to\Borel{S}$ for a measurable space $S$ as objects, a morphism $\oben{F}: \bigl(\Rational\to\Borel{S}\bigr) \to \bigl(\Rational\to\Borel{T}\bigr)$ is induced by a map $F: \Borel{S}\to\Borel{T}$ so that $\oben{F}(\gamma)(q) = F(\gamma(q))$ for the object $\gamma: \Rational\to\Borel{S}$  and $q\in\Rational$ holds. 
Denote by $\BQ$ the functor $\M\to\NQ$ which maps the measurable space $S$ to 
$
\{\gamma\mid \gamma: \Rational\to\Borel{S} \text{ is a map}\},
$ 
and $f: S\to T$ measurable is mapped to $\oben{f^{-1}}$, thus $\BQ$ is contravariant. 

Assume that $\tau: \Rational\times\B\nPfeil\B$ is a natural transformation, thus 
$
\tau_S(\cdot, A): q \mapsto \tau_S(q, A)\in\Borel{S}
$
is an object on $\NQ$ for each measurable space $S$ and for each $A\in\Borel{S}$. 

\BeginLemma{nat-transf-hat}
Put 
\begin{equation*}
\oben{\tau_S}(A) := \tau_S(\cdot, A)
\end{equation*}
for a natural transformation $\tau: \Rational\times\B\nPfeil\B$ and $A\in\Borel{S}$, then $\oben{\tau}: \B\nPfeil\BQ$ is a natural transformation. 
\EndLemma

\BeginProof
In fact, if $f: S \to T$ is a measurable map, then we have for the measurable set $A\in\Borel{S}$ and $q\in\Rational$
\begin{align*}
\oben{\tau_S}(\B(f)(A))(q)
& = 
\tau_S(q, \InvBild{f}{A})\\
& = 
(\tau_S\circ (\Rational\times\B)(f))(q, A)\\
& = 
\B(f)(\tau_T(q, A))\\
& =
\BQ(f)(\oben{\tau_T}(A))(q).
\end{align*}
\EndProof

\BeginCorollary{nat-transf-hat-rel}
$\oben{\tau}: \BS\nPfeil\BQ$ is a natural transformation, provided $\tau: \Rational\times\BS\nPfeil\B$ is natural.
\QED
\EndCorollary

As an illustration, each stochastic relation induces a natural transformation $\Rational\times\BS\nPfeil\B$ via the evaluation map.
 
\BeginLemma{SR-isNat}
Let $K: S \Trans S$ be a stochastic relation. Then 
\begin{equation*}
\varpi_K(q)(A) := \{s \in S \mid K(s)(A) < q\}
\end{equation*}
defines a natural transformation $\varpi: \RationalS\times\BS\nPfeil\BS$.
\EndLemma

\BeginProof
Because 
$
\varpi_K(q)(A) = \InvBild{K}{\basB_{q, A}},
$
and since $K$ is a measurable map, we infer  $\varpi_K(q)(A)\in\Borel{S}$, whenever $K: S \Trans S$. Now let $f: K \to L$ be a morphism, and take  $\langle q, B\rangle \in \Rational\times \Borel{T}$, then 
\begin{align*}
\bigl(\B(f)\circ \varpi_L\bigr)(q, B)
&=
\InvBild{f}{\{t \in T \mid L(t)(B) < q\}}\\
& = 
\{s\in S \mid K(s)(\InvBild{f}{B}) < q\}\\
& =
\bigl(\varpi_K\circ\Rational\times\BS(f)\bigr)(q, B).
\end{align*}
\EndProof

Another consequence is interesting for us as well.
\BeginCorollary{nat-transf-hat-phi}
Assume that 
$
\Phi: (\BQ)^I \nPfeil \BQ
$
is a natural transformation with $I = \{1, \dots, n\}$ for $n\in\Nat$ or $I = \Nat$ and that $\psi_i: \Rational\times\B\nPfeil\B$ for $i\in I$. Then
$
\oben{\Phi\bigl((\oben{\psi}_i)_{i\in I}\bigr)} 
$
defines a natural transformation $\oben{\Phi}:\Rational\times\B\nPfeil\B$ with
$
\oben{\Phi}_S(q)(A) = \Phi\bigl((\psi_{i, S}(\cdot, A))_{i\in I}\bigr)(q). 
$
\QED
\EndCorollary

To illustrate, define for rational $q > 0$ the sets
\begin{align*}
Q^{(n)}(q) 
& :=
\{a \in \Rational^n \mid a_1 + \dots + a_n \leq q\}\\
Q^{(\infty)}(q)
& := 
\{\Folge{a} \in \Rational^{\Nat_0} \mid a_0 + a_2 \dots \leq q\}
\end{align*}

\BeginExample{union-expl}
Let  $\langle \eta_1, \eta_2\rangle \in \BQ(S)\times\BQ(S)$  for a measurable space $S$, and define for $q\in\Rational$
\begin{equation*}
\Phi_S(\eta_1, \eta_2)(q) := \bigcup_{\langle a_1, a_2\rangle\in Q^{(2)}(q)}\bigl(\eta_{1, S}(a_1)\cap\eta_{2, S}(a_2)\bigr) 
\end{equation*}
Then $\Phi: \BQ\times\BQ\nPfeil\BQ$ is a natural transformation.

In fact, because 
$
\eta_1(a_1), \eta_2(a_2)\in\Borel{S}
$
for $\langle a_1, a_2\rangle\in Q^{(2)}(q)$, and because $Q^{(2)}(q)$ is countable, we infer that $\Phi_S(\eta_1, \eta_2)\in\BQ(S)$. Now let $f: S \to T$ be a measurable map, then this diagram commutes:
\begin{equation*}
\xymatrix{
\bigl(\BQ\times\BQ\bigr)(T)\ar[rr]^{\Phi_T}\ar[d]_{\bigl(\BQ\times\BQ\bigr)(f)}&&\BQ(T)\ar[d]^{\BQ(f)}\\
\bigl(\BQ\times\BQ\bigr)(S)\ar[rr]^{\Phi_S}&&\BQ(S)
}
\end{equation*}
This is so since we have for $\langle \eta_1, \eta_2\rangle\in\bigl(\BQ\times\BQ\bigr)(T)$
\begin{align*}
\Phi_S\bigl(\BQ(f)(\eta_1), \BQ(f)(\eta_2)\bigr)(q)
&=
\bigcup_a \bigl(\InvBild{f}{\eta_1(a_1)}\cap\InvBild{f}{\eta_2(a_2)}\bigr)\\
&= 
f^{-1}\bigl[\bigcup_a(\eta_1(a_1)\cap\eta_2(a_2))\bigr]\\
&=
\BQ(f)\bigl(\Phi_T(\eta_1, \eta_2)\bigr)
\end{align*}
\EndExample

The next example requires that the base spaces are closed under the Souslin operation.
\BeginExample{star-expl}
Let $\boldsymbol{\eta} := (\eta)_{n\in\Nat_0}\in \BR(S)^{\Nat_0}$, and define 
\begin{equation*}
\Psi_S(\boldsymbol{\eta})(q) := \bigcup\ \bigl\{\bigcap_{n\in\Nat_0} \eta_{n, S}(a_n)\mid a\in Q^{(\infty)}(q)\bigr\}
\end{equation*}
for $q\in\Rational$. Then $\Psi: (\BQ)^{\Nat_0} \nPfeil \BQ$, when functor $\B$ is restricted to category $\V$. 

We show first that $\Psi_S(\boldsymbol{\eta})(q)\in\Borel{S}$ whenever $S$ is closed under the Souslin operation. For this, we construct for $q > 0$ rational a bijection $\xi: \Nat_0^\Nat\to Q^{(\infty)}(q)$ such that $\nu|n = \nu'|n$ implies $\xi(\nu)|n = \xi(\nu')|n$ for all $\nu, \nu'\in\Nat_0^\Nat$ and all $n\in\Nat$, see~\cite[Lemma 4.6]{EED-PDL-TR}. We infer in particular that $\nu|n = \nu'|n$ implies $\xi(\nu)_n = \xi(\nu')_n$ for all $n\in\Nat$. Now put  
$
C_{\nu|n} := \eta_n\bigl(\xi(\nu)_n\bigr) \in \Borel{S}, 
$
then 
\begin{equation*}
\Psi_S(\boldsymbol{\eta})(q) = \bigcup_{\nu\in\Nat_0^\Nat} \bigcap_{n\in\Nat} C_{\nu|n}.
\end{equation*}
Since $S$ is closed under the Souslin operation, the assertion on measurability follows. Naturalness is then shown exactly as in Example~\ref{union-expl}. 
\EndExample

\section{Interpretations}
\label{sec:interpr}

We now turn to interpretations for PDL ---~although we did not define PDL yet, but never mind. A Kripke model will be employed for interpreting each simple program, similarly, an interpretation for primitive statements will be provided. We will build up from these data an interpretation for modal formulas in which the modalities are given through programs. This will be done through the Kleisli composition for the underlying monad, yielding an interpretation of basic blocks, i.e., of runs of simple programs, and through the natural transformations which will be associated with composing programs through nondeterministic choice and indefinite iteration. It will be convenient separating these notions, so we will first define what a Kripke model is, and then define models by adding these transformations. Morphisms will be important as well. They are defined for Kripke models, and, since the transformations for the complex program operations are natural, they carry over in a most natural fashion to models.

\subsection{Kripke Models}
\label{sec:kripke-models}

A \emph{stochastic Kripke model} $\K = (S, (K_\varrho)_{\varrho\in\Ur}, V)$ is a measurable space $S$ together with a family $(K_\varrho)_{\gamma\in\Ur}$ of stochastic relations $K_\gamma: S \Trans S$ such that
\begin{itemize}
\item $K_\epsilon = 1_S$,
\item $V: \P \to\Borel{S}$ is a map.
\end{itemize}
Here $1_S: S \Trans S$ is the identity relation
\begin{equation*}
1_S(s)(A) := 
\begin{cases}
1, & \text{if }s \in A \\
0, & \text{otherwise}.
\end{cases}
\end{equation*}
The set $V(p)$ gives for the atomic proposition $p\in\P$ the set of all states in which $p$ is assumed to hold. 

Given a primitive program $\varrho\in\Ur$, the stochastic relation $K_\varrho$ governs the transition upon executing $\varrho$: the probability that after executing program $\gamma$ in state $s\in S$ we are in a state which is an element of $A\in\Borel{S}$ is given by $K_\gamma(s)(A)$. Note that $K_\gamma(s)(S) < 1$ is not excluded, accounting for nonterminating programs.

A morphism of Kripke models is compatible with the transition structure for each simple program, and it respects the interpretation for primitive statements, formally:

\BeginDefinition{Kripke-morph}
Given Kripke models $\K = (S, (K_\varrho)_{\varrho\in\Ur}, V)$ and $\L = (T, (L_{\varrho})_{\varrho\in\Ur}, W)$, a measurable map $f: S \to T$ is a morphism $\K\to\L$ for the Kleisli models iff 
\begin{enumerate}
\item $f: K_{\varrho}\to L_{\varrho}$ is a morphism of stochastic relations for each $\varrho\in\Ur$,
\item $\InvBild{f}{W(p)} = V(p)$ for each atomic proposition $p\in\P$. 
\end{enumerate}
\EndDefinition
Thus for morphism $f: \K\to\L$ an atomic proposition $p$ holds in state $s$ iff it holds in $f(s)$, and the probability of hitting a state in $B\in\Borel{T}$ after executing program $\varrho$ in state $f(s)$ is the same as the probability of hitting a state in $\InvBild{f}{B}$ after executing $\varrho$ in state $s$.

We will need later that Kripke models are closed under coproducts, hence we state as an example the corresponding construction.
\BeginExample{summe}
Given Kripke models $\K = (S, (K_\varrho)_{\varrho\in\Ur}, V)$ and $\L = (T, (L_{\varrho})_{\varrho\in\Ur}, W)$, define the sum $\K\oplus\L$ of $\K$ and $\L$ as the Kripke model 
\begin{equation*}
\K\oplus\L := (S + T, ((K+L)_\varrho)_{\varrho\in\Ur}, V + W).
\end{equation*}
Here the measurable space $S + T$ carries the final $\sigma$-algebra with respect to the embeddings $i_S$ and $i_T$, and 
$
(K+L)_\varrho: (S + T) \Trans (S + T)
$
is defined through  
\begin{equation*}
(K+L)_\varrho (z)(A) := 
\begin{cases}
K_\varrho(s)(\InvBild{i_S}{A})& \text{ if } z = i_S(s), \\
L_\varrho(t)(\InvBild{i_T}{A})& \text{ if } z = i_T(t).
\end{cases}
\end{equation*}
Then 
$
\K\stackrel{i_S}{\longrightarrow} \K\oplus\L \stackrel{i_T}{\longleftarrow}\L
$
are morphisms. It is easy to see that $\K\oplus\L$ together with the embeddings is the coproduct. 
\EndExample 

Given a Kripke model $\K = (S, (K_\varrho)_{\varrho\in\Ur}, V)$, extend the transition laws from primitive programs to basic blocks, i.e., sequences of primitive programs upon setting
\begin{equation}
\label{basic-blocks}
K_{\varrho_1; \dots; \varrho_n} := K_{\varrho_1} \klComp \dots \klComp K_{\varrho_n},
\end{equation}
where for $K_i: S \Trans S$ $(i = 1, 2)$ the \emph{Kleisli composition} $K_1\klComp K_2$ of $K_1$ and $K_2$ is defined through 
\begin{equation*}
\bigl(K_1\klComp K_2\bigr)(s)(A) := \int_S K_2(t)(A)\ K_1(s)(dt)
\end{equation*}
($s\in S, A \in \Borel{S}$), see~\cite{Giry}; this operation is known as the \emph{convolution} of two transition kernels in probability theory. Interpreting equation (\ref{basic-blocks}) for two programs $\varrho_1, \varrho_2\in\Ur$, we see that after executing $\varrho_1$ in state $s$ the system goes into some intermediate state $t\in S$ from which program $\varrho_2$ continues, giving the probability of ending up in a state in Borel set $A$ as 
$
K_{\varrho_2}(t)(A).
$
The intermediate states are averaged over through $K_{\varrho_1}(s)$, accounting for the probability
\begin{equation*}
\int_S K_{\varrho_2}(t)(A)\ K_{\varrho_1}(s)(dt),
\end{equation*} 
which is just
$
\bigl(K_{\varrho_1}\klComp K_{\varrho_2}\bigr)(s)(A).
$

Notice that 
\begin{equation*}
K_\epsilon\klComp K_\varrho = K_\varrho = K_\varrho\klComp K_\epsilon
\end{equation*}
for all $\varrho\in\Ur$. Because stochastic relations are Kleisli morphisms for a monad, hence morphisms in a category, it follows that Kleisli composition is associative, thus we record for later use that
\begin{equation}
\label{wgKlComp}
(K_1\klComp K_2)\klComp K_3 = K_1\klComp(K_2\klComp K_3)
\end{equation}
holds (which we have already silently made use of in equation (\ref{basic-blocks})). 

\medskip

This extension from $\Ur$ to $\wrd{\Ur}$ through Kleisli composition is compatible with morphisms.
\BeginLemma{ext-morph}
Let $f: K_1 \to L_1$ and $f: K_2 \to L_2$ be morphisms of stochastic relations for $K_i: S \Trans S$ and $L_i: T \Trans T$ ($i = 1, 2$). Then $f: K_1\klComp K_2 \to L_1\klComp L_2$ is a morphism. 
\EndLemma

\BeginProof
This follows from Lemma~\ref{CVT}: 
\begin{align*}
\bigl(L_1\klComp L_2\bigr)(f(s))(B)
&=
\int_T L_2(y)(B)\ L_1(f(s))(dy) \\
&=
\int_T L_2(y)(B)\ \big(\SubProb{f}\bigl(K_1(s)\bigr)\big)(dy)\\
&=
\int_S L_2(f(x))(B)\ K_1(s)(dx) \\
&=
\int_S K_2(x)(\InvBild{f}{B})\ K_1(s)(dx)\\
&=
\bigl(K_2\klComp K_1)(s)(\InvBild{f}{B})\\
&=
\bigl(\SubProb{f}\circ (K_1\klComp K_2)\bigr)(s)(B).
\end{align*}
\EndProof

Applying this to morphisms for stochastic Kripke models yields

\BeginCorollary{cor-ext-morph}
Let $\K$ and $\L$ be Kripke models, and assume that $f: \K\to\L$ is a morphism. Then 
\begin{equation*}
f: K_{\varrho_1;\dots;\varrho_n}\to L_{\varrho_1;\dots;\varrho_n}
\end{equation*}
is a morphism for stochastic relations for all $\varrho_1;\dots;\varrho_n\in\wrd{\Ur}$.
\QED
\EndCorollary

Let $\Kr = \Kr(\Ur, \P)$ be the category of Kripke models with universally measurable state spaces; it has the morphisms according to the definition above. Hence the state space of an object in $\Kr$ is a measurable space which is closed under universal completion according to  Definition~\ref{completion}.  We define the functor $\R: \Kr\to\N$ from Kripke models to Borel sets of measurable spaces by adapting the Borel functor to $\Kr$: each Kripke model $\K = \bigl(S, (K_\varrho)_{\varrho\in\Ur}, V\bigr)$ is mapped to $\B(S)$. By the choice of the base category of universally measurable spaces we make sure that $\R(\K)$ is always closed under the Souslin operation. A morphism $f: \K\to\L$ is mapped by $\R$ to $f^{-1}: \Borel{T}\to\Borel{S}$. 

\def\B{\R}
  
Assume furthermore that we are given natural transformations $\Phi: \BQ\times\BQ \nPfeil \BQ$ and $\Psi: (\BQ)^\Nat \nPfeil \BQ$. We associate with each basic block $\varrho_1; \dots; \varrho_n$ a natural transformation 
$
\Gamma(\varrho_1; \dots; \varrho_n): \Rational\times \B\nPfeil \B
$
upon setting
\begin{equation}
\label{b-blocks}
\Gamma(\varrho_1; \dots; \varrho_n) := \varpi_{K_{\varrho_1; \dots; \varrho_n}}. 
\end{equation}
Assume that we have defined natural transformations 
$
\Gamma(\beta_1), \Gamma(\beta_2)
$
for the irreducible programs $\beta_1, \beta_2\in\ProIrr$, then
\begin{equation}
\label{b-union}
\Gamma(\beta_1\cup\beta_2) := \oben{\Phi(\oben{\Gamma(\beta_1)}, \oben{\Gamma(\beta_2)})}
\end{equation} 
defines a natural transformation
$
\Gamma(\beta_1\cup\beta_2): \Rational\times\B\nPfeil\B.
$
If 
$
\Gamma(\beta_n): \Rational\times \B\nPfeil \B
$
is defined for $\beta_n\in\ProIrr$, define
\begin{equation}
\label{b-star}
\Gamma\bigl(\infOp\langle\beta_n|n\in\Nat_0\rangle\bigr) := \oben{\Psi\bigl((\oben{\Gamma(\beta_n)})_{n\in\Nat_0}\bigr)}.
\end{equation}
Then 
$
\Gamma\bigl(\infOp\langle\beta_n|n\in\Nat_0\rangle\bigr): \Rational\times \B\nPfeil \B.
$

Summarizing, we note for the record
\BeginProposition{s-nat-transf}
Given the transformations $\Phi$ and $\Psi$ as above, $\Gamma(\beta): \Rational\times\B\nPfeil\B$ is a natural transformation, whenever $\beta$ is an irreducible program. 
\QED
\EndProposition

\medskip

It is worth noting that
\begin{itemize}
\item composition of programs is modelled through the composition operator for stochastic relations, hence through Kleisli composition for the underlying monad; this is the basic mechanism which the other transformations start from,
\item once a natural transformation for each basic block in $\wrd{\Ur}$ is defined, the Kripke model proper is only needed to give the semantics for the atomic propositions in $\P$. The transformations for irreducible programs $\beta_1\cup\beta_2$ and $\infOp\langle\beta_k|k\in\Nat\rangle$ now rests on the shoulders of the transformations $\Phi$ resp. $\Psi$.  
\end{itemize}

\subsection{Defining a Model}
\label{ref:model}

Now that the basic ingredients for defining a model are in place, we have to have a closer look at these components. It does not make sense to define a models with arbitrary transformations, because it is clear that the transformations should satisfy some properties, monotonicity and compatibility among that. The latter property refers to the observation that nondeterministic choice and indefinite iteration are somewhat related (this is reflected in the rewrite rule $\Star$), consequently we require their interpretations to cooperate along these lines. Some properties are captured in the definition below. 

\BeginDefinition{eig-opns}
Let $\Phi: \BQ\times\BQ \nPfeil \BQ$ and $\Psi: (\BQ)^{\Nat_0} \nPfeil \BQ$ be natural transformations.
\begin{enumerate}
\item\label{eig-opns-a} $\Phi$ is called
\begin{itemize}
\item \emph{associative}, iff 
$
\Phi\bigl(\eta_1, \Phi(\eta_2, \eta_3)\bigr) = \Phi\bigl(\Phi(\eta_1, \eta_2), \eta_3\bigr) 
$
\item \emph{commutative}, iff
$
\Phi(\eta_1, \eta_2) = \Phi(\eta_2, \eta_1),
$
\item \emph{idempotent}, iff
$
\Phi(\eta_1, \eta_1) = \eta_1,
$
provided $\eta_1$ is monotone (i.e., $q \mapsto \eta_{1, S}(q)(A)$ is a monotone map for each $A\in\Borel{S}$)
\end{itemize}
for any $\eta_1, \eta_2, \eta_3: \BQ \nPfeil \BQ$ holds.
\item\label{eig-opns-b} $\Psi$ is called \emph{symmetric} iff
\begin{equation*}
\Psi\bigl(\Psi((\eta_{i, j})_{i\in \Nat_0})_{j\in\Nat_0}\bigr)
=
\Psi\bigl(\Psi((\eta_{i, j})_{j\in \Nat_0})_{i\in\Nat_0}\bigr)
\end{equation*}
for each double indexed sequence $(\eta_{i, j})_{\langle i, j\rangle\in\Nat_0\times\Nat_0}$ with $\eta_{i, j}: \BQ \nPfeil \BQ$
for all $i, j\in\Nat_0$ holds.
\item\label{eig-opns-c} 
$\Phi$ and $\Psi$ are said to be \emph{compatible} iff
\begin{equation*}
\Psi((\eta_i)_{i\in\Nat_0}) = \Phi\bigl(\eta_0, \Psi((\eta_{i+1})_{i\in\Nat_0})\bigr)
\end{equation*}
holds for each sequence $(\eta_i)_{i\in\Nat_0}$ with $\eta_i: \BQ \nPfeil \BQ$ for each $i\in\Nat_0$. 
\end{enumerate}
\EndDefinition

The properties of $\Phi$ described in Definition~\ref{eig-opns} under $\ref{eig-opns-a}.$ make the set of all natural transformations $\BQ \nPfeil \BQ$ a commutative semigroup, if $\langle \eta_1, \eta_2\rangle$ is sent to $\Phi(\eta_1, \eta_2)$. They are modelled after union or intersection in the power set of a set. Property $\ref{eig-opns-b}.$ deals with evaluating operator $\Psi$: An infinite matrix of natural transformations may be evaluated first along its rows, producing a sequence of natural transformations again; evaluating this is assumed to be identical to evaluating the matrix first along the columns and then evaluating the results. Finally, property~$\ref{eig-opns-c}.$ says that $\Psi$ may be evaluated stepwise through operator $\Phi$ akin to an infinite sum, an infinite union, or an indefinite iteration.  
 
\BeginLemma{opns-have-prop}
The operators $\Phi$ and $\Psi$ defined in Example~\ref{union-expl} resp. Example~\ref{star-expl} have these properties:
\begin{enumerate}[a.]
\item\label{opns-have-prop-a} $\Phi$ is associative, commutative and idempotent,
\item\label{opns-have-prop-b} $\Psi$ is symmetric,
\item\label{opns-have-prop-c} $\Phi$ and $\Psi$ are  compatible.
\end{enumerate}
\EndLemma

\BeginProof
1.
Properties $\ref{opns-have-prop-a}$ and $\ref{opns-have-prop-c}$ are fairly obvious. Let $(\eta_{i, j})_{\langle i, j\rangle\in\Nat_0\times\Nat_0}$ with $\eta_{i, j}: \BQ \nPfeil \BQ$, put 
$
\boldsymbol{\rho}_i := (\eta_{i, j})_{j\in\Nat_0}
$
and
$
\boldsymbol{\sigma}_j := (\eta_{i, j})_{i\in\Nat_0}.
$
We now show that 
\begin{equation*}
\Psi_\K\bigl((A_i)_i\bigr) = \Psi_\K\bigl((B_j)_j\bigr)
\end{equation*}
holds, where
\begin{align*}
A_i(q) 
&=
\Psi_\K(\boldsymbol{\rho}_i)(q)\\
B_j(q) 
&=
\Psi_\K(\boldsymbol{\sigma}_j)(q)\\
\end{align*}
This will establish that operator $\Psi$ is symmetric.

2.
Now fix $q\in\Rational$ and put for $\Folge{a}\in\Q^{(\infty)}$
\begin{align*}
Z(a) 
& :=
\{(a_{i, j}) \mid \forall i\in \Nat_0: \sum_{j\in\Nat_0} a_{i, j} \leq a_i\},\\
R(a) 
& :=
\{(a_{i, j}) \mid \forall j\in \Nat_0: \sum_{i\in\Nat_0} a_{i, j} \leq a_j\}.\\
\end{align*}
Hence an infinite matrix of non negative numbers is in $Z(a)$ iff for each row $i$ the column sums are dominated by $a_i$, similarly for $R(a)$ and the row sums. Note that
\begin{equation}
\label{Prings}
\sum_{i\in\Nat_0}\bigl(\sum_{j\in\Nat_0} a_{i, j}\bigr) = \sum_{j\in\Nat_0}\bigl(\sum_{i\in\Nat_0} a_{i, j}\bigr) 
\end{equation}
by Pringsheim's Theorem~\cite[V.31]{Bromwich}, because all terms are non-negative.

3.
Now
\begin{align}
s \in \Psi_\K\bigl((A_i)_i\bigr)(q)
&\Longleftrightarrow\label{Prings-1}
\exists a \in Q^{(\infty)}(q)\forall i \in \Nat_0 \exists (a_{i, j})_j\in \Q^{(\infty)}(a_i)\forall j \in \Nat_0:
s \in \eta_{i, j, \K}(a_{i, j})\\
&\Longleftrightarrow\label{Prings-2}
\exists a \in Q^{(\infty)}(q)\exists b \in Z(a) \forall i, j\in \Nat_0: s \in \eta_{i, j, \K}(b_{i, j})\\
&\Longleftrightarrow\label{Prings-3}
\exists x \in Q^{(\infty)}(q)\exists y \in R(x) \forall i, j\in \Nat_0: s \in \eta_{i, j, \K}(y_{i, j})\\
&\Longleftrightarrow\label{Prings-4}
s \in \Psi_\K\bigl((B_j)_j\bigr)(q)
\end{align}
For, assume that $a$ and $b$ are given according to~(\ref{Prings-2}), then define 
$
x_j := \sum_{i\in\Nat_0} b_{i, j}, y := b,
$
hence 
\begin{equation*}
\sum_j x_j = \sum_j \sum_i b_{i, j} \stackrel{(\ref{Prings})}{=} \sum_i \sum_j b_{i, j} \leq \sum_i a_i \leq q.
\end{equation*}
This justifies the implication $(\ref{Prings-2})\Rightarrow(\ref{Prings-3})$, similarly for the converse. 
\EndProof

\medskip

Call a natural transformation $\Lambda: (\BQ)^I\nPfeil\BQ$ \emph{monotone} iff $\Lambda\bigl((\eta_i)_{i \in I}\bigr)$ is monotone, provided  $\eta_i: \BQ\nPfeil\BQ$ is monotone for all $i \in I \subseteq \Nat_0$, see Definition~\ref{eig-opns}. 

We extend Kripke models now to models for PDL.

\BeginDefinition{Mod-PDL}
A \emph{model $\Mo = (\K, \Phi, \Psi)$ for PDL} is composed of a Kripke model $\K$ and of two monotone transformations $\Phi: \BQ\times\BQ \nPfeil \BQ$ and $\Psi: (\BQ)^{\Nat_0} \nPfeil \BQ$ so that $\Phi$ is associative, commutative and idempotent, $\Psi$ is symmetric, and $\Phi$ and $\Psi$ are compatible. 
\EndDefinition

When talking about a model, we always refer to a model in the sense of Definition~\ref{Mod-PDL}, unless otherwise specified. Hence we always have with a model a Kripke model and two transformations at our disposal. Define for model $\Mo$ the transformation $\GM(\beta): \Rational\times\B\nPfeil\B$ for  irreducible programs $\beta$ as at the end of Section~\ref{sec:kripke-models}, equations~\ref{b-blocks} through~\ref{b-star}, see Proposition~\ref{s-nat-transf}.

\BeginLemma{is-monotone}
Given an irreducible program $\beta$, the  state space $S$ of a Kripke model $\K$, the map 
$
q \mapsto \GM[\Mo, \K](q, A) := \bigl(\GM(\beta)\bigr)_\K(q, A)
$
is monotone for any fixed $A \in \Borel{S}$.
\EndLemma

\BeginProof
This is established by induction on $\beta$. Assume first that 
$
\beta = \varrho_1;\dots;\varrho_n\in\wrd{\Ur}.
$
Then 
\begin{equation*}
\GM[\Mo, \K](\varrho_1;\dots;\varrho_n)(q, A) 
=
\{s \in S \mid K_{\varrho_1;\dots;\varrho_n}(s)(A) < q\},
\end{equation*}
which is clearly a monotone function of $q$. If $\beta = \beta_1\cup\beta_2$, and monotonicity is established already for $\beta_1$ and $\beta_2$, then $\oben{\GM(\beta_1)}$ and $\oben{\GM(\beta_2)}$ are monotone, thus 
$
\Phi(\oben{\GM(\beta_1)},\oben{\GM(\beta_2)})
$
is monotone, from which the assertion for $\beta$ follows. One argues similarly for 
$
\beta = \infOp\langle \beta_n | n \geq 0\rangle,
$
provided the claim holds for all $\beta_n$.
\EndProof

\medskip

We show now that $\GM$ is invariant under the equivalence classes with respect to $\equiv$, as far as irreducible programs are concerned. This step is necessary for ensuring that the interpretation of formulas is well defined. 

\BeginProposition{subs-kern}
Let $\beta_1, \beta_2$ be irreducible programs with $\beta_1 \equiv \beta_2$. Then $\GM(\beta_1) = \GM(\beta_2)$.
\EndProposition

\BeginProof
1.
It is enough to show that 
$
\beta_1\approx\beta_2
$
implies 
$
\GM(\beta_1) = \GM(\beta_2).
$
Because no rewrite rules apply due to irreducibility, we may then conclude that 
\begin{equation*}
\equiv\ \cap\ \bigl(\ProIrr\times\ProIrr\bigr) \subseteq \Kern{\GM},
\end{equation*}
from which the assertion follows. We will discuss the different cases in turn.

2.
The cases $\IdEpsl$ and $\IdEpsr$ are covered by the observation that $K_\epsilon$ = $1_S$, which in turn is the neutral element for Kleisli composition, case $\AssS$ follows from associativity for Kleisli composition. Because $\Phi$ is associative and commutative, the cases $\AssU$ resp. $\Comm$ are covered as well. We infer from Lemma~\ref{is-monotone} and from idempotence of $\Phi$ that $\GM(\beta_1\cup\beta_1) = \GM(\beta_1)$. Finally, the cases $\InfDisj$ and $\Transp$ are covered through the compatibility of $\Phi$ and $\Psi$ resp. the symmetry of $\Psi$. 
\EndProof
 
Now take a program $\pi\in\Pro$ and consider $\beta_1, \beta_2\in\Theta(\pi) \cap \ProIrr$. Then $\GM(\beta_1) = \GM(\beta_2)$. Sending $\Theta(\pi) \cap \ProIrr$ to $\GM(\beta)$, provided $\beta\in\Theta(\pi) \cap \ProIrr$, we obtain a well defined map (recall $\Theta(\pi) \cap \ProIrr \not= \emptyset$ by Corollary~\ref{each-irred}). 

Thus define
\begin{equation}
\label{def_i}
\It_\Mo(\pi) := \GM(\beta),
\end{equation}
with $\pi\in\Pro$, provided $\beta\in\Theta(\pi)\cap\ProIrr$. This is defines a natural transformation, see Proposition~\ref{s-nat-transf}.

\section{The Logics}
\label{sec:theLogic}

We define the logic PDL as usual through modal operators which come from programs; because we investigate probabilistic aspects, we introduce a quantitative aspect by limiting certain probabilities from above. The logic is negation free and does not have disjunction. This looks on first sight a bit restricting, but since we work in a Boolean algebra of sets we can express negation through complementation, hence we do not need a separate operator for it. Omission of disjunction, however, cannot be compensated; it turns out that disjunction is not really necessary in the arguments to follow, so Occam's Razor could be applied. It should also be noted that we do not include the test operator. While this operator expands the usability of the logic, it does not contribute to the structural questions which we are concerned with; this has been discussed in~\cite[Section 6.5]{EED-PDL-TR}.

We will first define PDL and its semantics, then we will take only the simple programs and the atomic expressions and define a Hennessy-Milner logic from it, much in the spirit of~\cite{Larsen+Skou,Desharnais-Edalat-Panangaden,EED-Book}. This type of logics has been investigated extensively, and it will be helpful to use its semantic properties for the investigation of PDL. Syntactically, we have in the Hennessy-Milner logic only basic blocks at our disposal, these basic blocks are important for expressing the semantics of programs in PDL, so that we will relate these constructs to each other.

Finally we define expressivity ---~logical equivalence, bisimilarity, behavioral equivalence~ÐÐÐ for our models and relate them to each other. Bisimilarity will play a special r{\^o}le which partly will have to be delegated to the next section due to Standard Borel spaces being closed under the Souslin operation only in the finite case. The constructions to be undertaken will require some leg work for constructing the proper measurable spaces etc.

\subsection{PDL}
\label{sec:pdl}
Given a set $\Ur$ of primitive programs and a set $\P$ of atomic propositions, we define the formulas of logic $\PL$ through this grammar
\begin{equation*}
\phi ::= \top \ \mid\  p \ \mid\  \phi_1\wedge\phi_2\ \mid\  \md{\pi}{q}\phi
\end{equation*}
with $p\in\P$ an atomic proposition, $\pi\in\Pro$ a program and $q\in\Rational$ a rational number. Hence a formula is $\top$ as a formula which always holds, an atomic proposition, the conjunction of two formulas or a modal formula $\md{\phi}{q}\phi$. The latter one is going to hold whenever formula $\phi$ holds with probability less than $q\in\Rational$ after executing program $\pi$. 

Define inductively for a given model $\Mo = (\K, \Phi, \Psi)$ with state space $S$ and valuation $V: S \to \Borel{S}$  the \emph{extension} or \emph{validity set} $\Gilt_\Mo$ for formula $\phi$ through
\begin{align}
\label{val-a}\gilt{\top} & := S,\\
\label{val-b}\gilt{p} & := V(p),\\
\label{val-c}\gilt{\phi_1\wedge\phi_2} & := \gilt{\phi_1}\cap\gilt{\phi_2},\\
\label{val-d}\gilt{\md{\pi}{q}\phi} & := \It_\Mo(\pi)(q)(\gilt{\phi}),
\end{align}
where the natural transformation $\It_\Mo$ is defined in Equation~(\ref{def_i}). The validity relation $\models$ is then defined through
\begin{equation*}
\Mo, s \models \phi \Longleftrightarrow s \in \gilt{\phi},
\end{equation*}
consequently, $\Mo, s \models \top$ holds by~$(\ref{val-a})$ always, and $\Mo, s \models p$ iff $s \in V(p)$ for the atomic proposition $p\in\P$  by~$(\ref{val-b})$. If $\varrho_1, \dots, \varrho_n\in\Ur$, we infer from~$(\ref{val-c})$ through the definition of $\It$ in particular
\begin{equation}
\label{m-holds-for-block}
\Mo, s \models \md{\varrho_1; \dots; \varrho_n}{q}\phi
\text{ iff }
K_{\varrho_1; \dots; \varrho_n}(s)(\gilt\phi) < q
\end{equation}
Although the logic is negation free, we are still able to state that formula $\phi$ \emph{does not hold} in a state. Because we work in a $\sigma$-algebra, thus in particular in a Boolean algebra, we can state that formula $\phi$ does not hold in state $s$ iff $s\not\in\gilt\phi$, so that the set
$
\{s \in S \mid \phi\text{ does not hold in } s\}
$
is a measurable set, provided the extension of $\phi$ is measurable. 

We note for later use that the validity sets are measurable. This is so since we deal with natural transformations involving the Borel functor. 

\BeginLemma{ext-meas}
$\gilt\phi\in\Borel{S}$ for a model $\Mo$ over state space $S$ and a PDL formula $\phi$.\QED
\EndLemma

\BeginExample{old-form}
Consider the transformations $\Phi$ from Example~\ref{union-expl} and $\Psi$ from Example~\ref{star-expl}. Expanding~(\ref{val-d}), we obtain
\begin{align}
\label{old-form-a}\gilt{\md{\pi_1\cup\pi_2}{q}\phi}
& =
\bigcup\bigl\{\gilt{\md{\pi_1}{a_1}\phi}\, \cap\, \gilt{\md{\pi_2}{a_2}\phi} \mid  a_1, a_2 \in\Rational, a_1 + a_2 \leq q\bigr\},\\
\label{old-form-b}\gilt{\md{\pi^*}{q}\phi}
&=
\bigcup\bigl\{\bigcap_{m\in\Nat_0}\gilt{\md{\pi^m}{a_m}\phi} \mid \Folge{a}\subseteq\Rational, \text{ for all } n \in \Nat_0, \sum_n a_n \leq q\bigr\}.
\end{align}

Selecting nondeterministically one of the programs $\pi_1$ or $\pi_2$, $\gilt{\md{\pi_1}{a_1}\phi}$ accounts for all states which are lead by executing $\pi_1$ to a state in which $\phi$ holds with probability at most $a_1$, similarly, $\gilt{\md{\pi_2}{a_2}\phi}$ for $\pi_2$. Since we want to bound the probability from above by $q$, we require $a_1 + a_2 \leq q$. This leads to Equation~(\ref{old-form-a}).

Suppose that executing program $\pi$ exactly $n$ times results in a state in which $\phi$ holds with probability not exceeding $a_n$, then executing $\pi$ a finite number of times (including not executing it at all) results in a member of $\gilt\phi$ with probability at most $a_0 + a_1 + \dots $, which should be bounded above by $q$ for the resulting state to be a state in which $\phi$ holds with probability at least $q$. This leads to Eq.~(\ref{old-form-b}).

These specific interpretations were investigated more closely in~\cite{EED-PDL-TR}. 
\EndExample

%
Define for each state $s$ of a model $\Mo$ the \emph{$\Mo$-theory associated with $s$} as the set of formulas which hold in that state, formally
\begin{equation*}
\theTheory{s}{\PL} := \{\phi \mid \phi \text{ is a formula in $\PL$ and } \Mo, s \models \phi\}.
\end{equation*}

\subsection{A simple Hennessy-Milner logic}
\label{sec:HM-logic} 
We define the negation free Hennessy-Milner logic $\PM$ through these formulas:
\begin{equation*}
\phi ::= \top\ \mid\  p\ \mid\  \phi_1\wedge\phi_2\ \mid\  \langle \varrho \rangle_q \phi
\end{equation*}
with $\varrho\in\Ur$ a primitive program, $q \in \Rational$ a threshold value, and $p\in\P$ an atomic proposition. Thus each primitive program serves as a modal operator of arity 1 for the modal logic $\PM$. 

\medskip

Considering $\varrho$ as an action as in labelled Markov transition systems, the intended interpretation of formula $\langle \varrho \rangle_q \phi$ holding in state $s$ is that upon action $\varrho$, i.e., upon executing program $\varrho\in\Ur$, a state in which $\phi$ holds is reached with probability at least $q$, see, e.g.~\cite{Larsen+Skou,Desharnais-Edalat-Panangaden,EED-Book}. 

Formally, we define for a Kripke model $\K = (S, (K_\varrho)_{\varrho\in\Ur}, V)$ and each formula $\phi$ of $\PM$ the validity sets $\kgilt\phi$ recursively through
\begin{align}
\label{k-val-a}\kgilt{\top} 
& :=
S,\\
\label{k-val-b}\kgilt{p} 
& := 
V(p), \text{ if } p \in \P,\\
\label{k-val-c}\kgilt{\phi_1\wedge\phi_2}
& :=
\kgilt{\phi_1}\cap\kgilt{\phi_2},\\
\label{k-val-d}\kgilt{\langle \varrho \rangle_q \phi} 
& :=
\{s \in S \mid K_\varrho(s)(\kgilt\phi) \geq q\}
\end{align} 
Define for state $s$ and formula $\phi$ the relation $\models$ through 
\begin{equation*}
\K, s \models \phi \Leftrightarrow s \in \kgilt\phi,
\end{equation*}
Equation~($\ref{k-val-d}$) shows that $\kgilt\phi$ is always a measurable set.  A comparison with $\gilt\cdot$ shows that the definitions for $\top$, for atomic propositions, and for the conjunction of formulas~($\ref{val-a},\ref{val-b}, \ref{val-c})$ resp.~($\ref{k-val-a},\ref{k-val-b}, \ref{k-val-c})$ are identical. Because of the identity~($\ref{m-holds-for-block}$), we see that for $\varrho\in\Ur$ and a formula $\phi$ which is both an $\PM$ and an $\PL$ formula the correspondence
\begin{equation}
\label{relate-uno}
\gilt{\md{\varrho}{q}\phi} = S \setminus \kgilt{\langle \varrho \rangle_q \phi}
\end{equation}
holds. This observation can be refined. Define
\begin{align*}
\IKa{A}{\varrho, q} 
& :=
\{s \in S \mid K_\varrho(s)(A) \geq q\},\\
\IK{A}{\varrho_1, q_1, \dots, \varrho_{n+1}, q_{n+1}}
& :=
\IKa{\IK{A}{\varrho_1, q_1, \dots, \varrho_n, q_n}}{\varrho_{n+1}, q_{n+1}}\\
\IMa{A}{\varrho, q} 
& :=
\{s \in S \mid K_\varrho(s)(A) < q\},\\
\IM{A}{\varrho_1, q_1, \dots, \varrho_{n+1}, q_{n+1}}
& :=
\IMa{\IM{A}{\varrho_1, q_1, \dots, \varrho_n, q_n}}{\varrho_{n+1}, q_{n+1}}.
\end{align*}
for the measurable set $A\in\Borel{S}$, $\varrho, \varrho_1, \dots, \varrho_n, \varrho_{n+1}\in\Ur$ and $q, q_1, \dots q_n, q_{n+1}\in\Rational$. Thus, e.g.,
\begin{align*}
\IK{\kgilt{p}}{\varrho_1, q_1, \varrho_2, q_2} 
& =
\kgilt{\langle \varrho_2\rangle_{q_2}\langle \varrho_1\rangle_{q_1}\ p}\\
\IM{\gilt{p}}{\varrho_1, q_1, \varrho_2, q_2} 
& =
\gilt{\md{\varrho_2}{q_2}\md{\varrho_1}{q_1}\ p}\\
\end{align*}
for the atomic program $p\in\P$. 

Note that $q \mapsto \IMa{A}{\varrho, q}$ is monotonically increasing, and that
$
\IK{A}{\varrho, q} = S \setminus \IM{A}{\varrho, q}
$
by Equation~$(\ref{relate-uno})$.

These quantities can be related for the probabilistic case.

\BeginLemma{strange-formulas}
Assume that $K_\varrho(s)(S) = 1$ for all states $s\in S$, then
\begin{multline}
\label{strange-formulas-a}
\IK{A}{\varrho_1, q_1, \dots, \varrho_{2\cdot n}, q_{2\cdot n}}
=\\
\bigcap \bigl\{
\IM{A}{\varrho_1, q_1, \varrho_2, 1-q_2+1/k_1, \varrho_3, q_3, \dots, \\\varrho_{2\cdot n}, 1-q_{2\cdot n}+1/k_n} \mid k_1, \dots, k_n \in \Nat
\bigl\}
\end{multline}
and
\begin{multline}
\label{strange-formulas-b}
\IK{A}{\varrho_1, q_1, \dots, \varrho_{2\cdot n+1}, q_{2\cdot n+1}}
=\\
\bigcap \bigl\{
S\setminus \IM{A}{\varrho_1, q_1, \varrho_2, 1-q_2+1/k_1, \varrho_3, q_3, \dots, \varrho_{2\cdot n}, 1-q_{2\cdot n}+1/k_n,\\\varrho_{2\cdot n+1}, q_{2\cdot n+1}} \mid k_1, \dots, k_n \in \Nat
\bigl\}
\end{multline}

\EndLemma

\BeginProof
The proof proceeds by induction on $n$. If $n = 0$, then there is nothing to prove for Equation~$(\ref{strange-formulas-a})$, and Equation~$(\ref{strange-formulas-b})$ boils down to
\begin{equation*}
\IK{A}{\varrho, q}
= 
\{s \in S \mid K_\varrho(s)(A) \geq q\}
 =
S \setminus \{s \in S \mid K_\varrho(s)(A) < q\}
 =
S \setminus \IM{A}{\varrho, q}.
\end{equation*}

Now assume that Equation~$(\ref{strange-formulas-a})$ and~$(\ref{strange-formulas-b})$ are established for $n$. 
Put 
\begin{align*}
T_{k_1, \dots, k_n} 
& := S \setminus R_{k_1, \dots, k_n},\\
R_{k_1, \dots, k_n} & := \IM{A}{\varrho_1, q_1, \varrho_2, 1-q_2+1/k_1, \\&\phantom{:=\IM{A}{\varrho_1, q_1,}}\varrho_3, q_3, \dots, \varrho_{2\cdot n}, 1-q_{2\cdot n}+1/k_n,\varrho_{2\cdot n+1}, q_{2\cdot n+1}},
\end{align*}
then
\begin{align*}
\IK{A}{\varrho_1, q_1, \dots, \varrho_{2\cdot n+1}, q_{2\cdot n+1}, &\ \varrho, q}\\
&
=
\IKa{\IK{A}{\varrho_1, q_1, \dots, \varrho_{2\cdot n+1}, q_{2\cdot n+1}}}{\varrho, q}\\
&
\stackrel{(*)}{=}
\{s \mid K_{\varrho}(s)(\bigcap_{k_1, \dots, k_n\in\Nat} T_{k_1, \dots, k_n}) \geq q\}\\
& 
=
S\setminus \{s \mid K_{\varrho}(s)(\bigcap_{k_1, \dots, k_n\in\Nat} T_{k_1, \dots, k_n}) < q\}\\
& 
\stackrel{(\sigma)}{=}
S\setminus \{s \mid \inf_{k_1, \dots, k_n\in\Nat} K_{\varrho}(s)(T_{k_1, \dots, k_n}) < q\}\\
& 
\stackrel{(p)}{=}
S\setminus \{s \mid 1 - \sup_{k_1, \dots, k_n\in\Nat} K_{\varrho}(s)(R_{k_1, \dots, k_n}) < q\}\\
& 
=
\{s \mid \sup_{k_1, \dots, k_n\in\Nat} K_{\varrho}(s)(R_{k_1, \dots, k_n}) \leq 1 - q\}\\
& 
=
\bigcap_{k_1, \dots, k_n\in\Nat}\{s \mid K_{\varrho}(s)(R_{k_1, \dots, k_n}) \leq 1 - q\}\\
& 
=
\bigcap_{k_1, \dots, k_n, k_{n+1}\in\Nat}\{s \mid K_{\varrho}(s)(R_{k_1, \dots, k_n})
 < 1 - q + 1/k_{n+1}\}\\
& 
=
\bigcap_{k_1, \dots, k_n, k_{n+1}\in\Nat}\IMa{R_{k_1, \dots, k_n}}{\varrho,1 - q + 1/k_{n+1}}.
\end{align*}

This implies Equation~$(\ref{strange-formulas-a})$ for $n+1$. The induction hypothesis is used in equality $(*)$, and equality $(\sigma)$ uses $\sigma$-additivity of the measure $K_{\varrho}(s)$ for each $s$: this property is equivalent to
\begin{equation*}
K_\varrho(s)\bigl(\bigcap_{n\in\Nat} A_n\bigr) = \inf_{n\in\Nat} K_\varrho(s)(A_n),
\end{equation*}
whenever $\Folge{A}\subseteq\Borel{S}$ is decreasing. 
Finally, equality $(p)$ uses the assumption that the full space has probability one. 

To work on Equation~$(\ref{strange-formulas-b})$ for $n+1$, put
\begin{equation*}
V_{k_1, \dots, k_{n+1}} := 
\IM{A}{\varrho_1, q_1, \varrho_2, 1 - q_2 + 1/k_1, \varrho_3, q_3, \dots, \varrho_{2\cdot(n+1)}, 1 - q_{2\cdot(n+1)} + 1/k_{n+1}},
\end{equation*}
then
\begin{align*}
\IK{A}{\varrho_1, q_1, \dots, \varrho_{2\cdot(n+1)}, q_{2\cdot(n+1)}, &\ \varrho, q}\\
&
=
\IK{\IK{A}{\varrho_1, q_1, \dots, \varrho_{2\cdot(n+1)}, q_{2\cdot(n+1)}}}{\varrho, q}\\
&
=
\{s \mid K_\varrho(\IK{A}{\varrho_1, q_1, \dots, \varrho_{2\cdot(n+1)}, q_{2\cdot(n+1)}}) \geq q\}\\
&
=
\{s \mid K_\varrho(\bigcap_{k_1, \dots, k_{n+1}\in\Nat} V_{k_1, \dots, k_{n+1}}) \geq q\}\\
&
=
\{s \mid \inf_{k_1, \dots, k_{n+1}\in\Nat} K_\varrho(V_{k_1, \dots, k_{n+1}}) \geq q\}\\
&
=
\bigcap_{k_1, \dots, k_{n+1}\in\Nat} \{s \mid K_\varrho(V_{k_1, \dots, k_{n+1}}) \geq q\}\\
&
=
\bigcap_{k_1, \dots, k_{n+1}\in\Nat} S\setminus\{s \mid K_\varrho(V_{k_1, \dots, k_{n+1}}) < q\}\\
&
=
\bigcap_{k_1, \dots, k_{n+1}\in\Nat} S\setminus \IM{V_{k_1, \dots, k_{n+1}}}{\varrho, q}
\end{align*}
Equation~$(\ref{strange-formulas-b})$ for $n+1$ follows now.
\EndProof

This has as a consequence that the semantics of a large class of formulas in $\PL$ can be expressed through the semantics for $\PM$-formulas.
\BeginCorollary{simple-expr}
Assume that $K_\varrho(s)(S) = 1$ for all states $s\in S$, and let $p$ be an atomic formula. Then
\begin{multline*}
\kgilt{\langle\varrho_{2\cdot n}\rangle_{q_{2\cdot n}}\dots\langle\varrho_1\rangle_{q_1} p}\\
=
\bigcap_{k_1, \dots, k_n\in\Nat}\gilt{\md{\varrho_{2\cdot n}}{1 - q_{2\cdot n} + 1/k_n}
\md{\varrho_{2\cdot n - 1}}{q_{2\cdot n - 1}}\dots \md{\varrho_{2}}{1 - q_{2} + 1/k_1}
\md{\varrho_{1}}{q_1}
\ p}
\end{multline*}
and
\begin{multline*}
\kgilt{\langle\varrho_{2\cdot n + 1}\rangle_{q_{2\cdot n + 1}}\dots\langle\varrho_1\rangle_{q_1} p}\\
=
\bigcap_{k_1, \dots, k_n\in\Nat}S\setminus \gilt{\md{\varrho_{2\cdot n + 1}}{q_{2\cdot n + 1}}\md{\varrho_{2\cdot n}}{1 - q_{2\cdot n} + 1/k_n}
\md{\varrho_{2\cdot n - 1}}{q_{2\cdot n - 1}}\dots\\ \md{\varrho_{2}}{1 - q_{2} + 1/k_1}
\md{\varrho_{1}}{q_1}
\ p}
\end{multline*}
\QED
\EndCorollary

Note that logic $\PM$ does not deal with the choice operator or with indefinite iteration ---~we do not even have disjunction in this logic after all. Hence we will not be able to interpret the semantics of these operators in $\PL$ through operators in $\PM$. 

\medskip

Returning to the general discussion, define as above
\begin{equation*}
\theTheory[\K]{s}{\PM} := \{\phi \mid \phi \text{ is a formula in $\PM$ and } \K, s \models \phi\}
\end{equation*}
the $\K$-theory associated with state $s$. 

It is not difficult to establish that validity is preserved under morphisms.

\BeginProposition{pres-val}
Let $\K_1$ and $\K_2$ be Kripke models, and $f: \K_1\to\K_2$ a morphism, then
\begin{equation*}
\K_1, s \models \phi \Longleftrightarrow \K_2, f(s) \models \phi
\end{equation*}
for each state $s$ in $\K_1$ and each $\PM$-formula $\phi$. 
\EndProposition
\BeginProof
See, e.g.,~\cite[Lemma 6.17]{EED-Book}.
\EndProof

\subsection{Expressivity}
\label{sec:expr}

Kripke models are traditionally related to each other in different ways, which are captured in the following definition. 
\medskip

\BeginDefinition{expr-kripke-mod}
Let $\K_1$ and $\K_2$ be Kripke models, then $\K_1$ and $\K_2$ are called
\begin{enumerate}
\item \emph{behaviorally equivalent} iff there exists a Kripke model $\K_0$ and surjective morphisms $f_1, f_2$ with 
$
\K_1 \stackrel{f_1}{\longrightarrow} \K_0 \stackrel{f_2}{\longleftarrow} \K_2,
$
\item \emph{HM-equivalent} iff 
$$
\{\theTheory[\K_1]{s}{\PM}\mid s \text{ is a state in }\K_1\}
=
\{\theTheory[\K_2]{t}{\PM}\mid t \text{ is a state in }\K_2\},
$$
\item \emph{bisimilar} iff there exists a Kripke model $\K_0$ and surjective morphisms $f_1, f_2$ with 
$$
\K_1 \stackrel{f_1}{\longleftarrow} \K_0 \stackrel{f_2}{\longrightarrow} \K_2.
$$
\end{enumerate}
\EndDefinition

The name \emph{HM-equivalence} alludes to the Hennessy-Milner logic which gives the context of this discussion. Usually the term ``logical equivalence'' is used. We will define logical equivalence below for models, and we do not want these very closely related concepts to get confused.   Thus $\K_1$ and $\K_2$ are behaviorally equivalent iff we can find an intermediate Kripke model which permits comparing the validity of formulas through surjective morphisms; we need surjectivity here because we want to be able to trace back a state in the intermediate Kripke model to $\K_1$ and $\K_2$. Otherwise we could simply take the coproduct of the Kripke models, see Example~\ref{summe}. The models are bisimilar iff we can find a mediating model for them, and they are HM equivalent iff we can find for each state in $\K_1$ another state in $\K_2$ which satisfies exactly the same formulas, and vice versa. The reader is referred to~\cite{Larsen+Skou,Desharnais-Edalat-Panangaden,EED-Book,EED-CS-Survey} for an extensive discussion stressing different angles.

Kripke models have been defined over the category of measurable spaces, the discussion of bisimilarity, however, requires some differentiation with respect to the base category for the state space. 

\medskip

The following result is well known.

\BeginTheorem{main-expr}
Let $\K_1$ and $\K_2$ be Kripke models, and consider these statements.
\begin{enumerate}[a.]
\item\label{main-expr-a} $\K_1$ and $\K_2$ are behaviorally equivalent.
\item\label{main-expr-b} $\K_1$ and $\K_2$ are HM-equivalent.
\item\label{main-expr-c} $\K_1$ and $\K_2$ are bisimilar.
\end{enumerate}
Then the following holds:
\begin{enumerate}[i.] 
\item\label{main-expr-i} $\ref{main-expr-a}. \Leftrightarrow \ref{main-expr-b}. \Leftarrow \ref{main-expr-c}$. 
\item\label{main-expr-ii} If $\K_1$ and $\K_2$ both are models over analytic spaces, and if both $\Ur$ and $\P$ are countable, then all three statements are equivalent. Moreover, if $\K_1$ and $\K_2$ are Kripke models over Polish spaces, then in this case a mediating model over a Polish space may be constructed. 
\end{enumerate}
\EndTheorem

\BeginProof
See~\cite[Theorem 6.17]{EED-CS-Survey} for~$\ref{main-expr-i}.$ and models over Polish spaces in~$\ref{main-expr-ii}$. The case of Kripke models over analytic spaces has first been discussed in~\cite{Edalat,Desharnais-Edalat-Panangaden}.
\EndProof

Snchez Perraf shows in~\cite{Sanchez} that the existence of a bisimulation is tied to analytic and, by implication, to Polish spaces. Hence an attempt to generalize part~$\ref{main-expr-ii}.$ of Theorem~\ref{main-expr} to general measurable spaces is futile. 

Given a model $\Mo = (\K, \Phi, \Psi)$, call $\K$ the  \emph{Kripke model underlying $\Mo$}. Define for models 
$
\Mo_1 = (\K_1, \Phi, \Psi)
$
and
$
\Mo_2 = (\K_2, \Phi, \Psi)
$
a \emph{model morphism} $f: \Mo_1\to\Mo_2$ as a morphism $f: \K_1\to\K_2$ for the underlying Kripke models. Note that $\Phi$ and $\Psi$ do not enter explicitly  into this definition because they are natural transformations, hence by their very nature compatible with morphisms for Kripke models. 

Behavioral equivalence and bisimilarity can be described in terms of these morphisms: 

\BeginDefinition{expr-mod}
Models $\Mo_1$ and $\Mo_2$ are \emph{behaviorally equivalent} iff there exists a model $\Mo_0$ and surjective morphisms $f_1, f_2$ with 
$
\Mo_1 \stackrel{f_1}{\longrightarrow} \Mo_0 \stackrel{f_2}{\longleftarrow} \Mo_2.
$
If  a mediating model $\Mo_3$ and surjective morphisms $g_1, g_2$ exist with 
$
\Mo_1 \stackrel{g_1}{\longleftarrow} \Mo_3 \stackrel{g_2}{\longrightarrow} \Mo_2,
$
then $\Mo_1$ and $\Mo_2$ are called \emph{bisimilar}. $\Mo_1$ and $\Mo_2$ are \emph{logically equivalent} iff 
$$
\{\theTheory[\Mo_1]{s}{\PL} \mid s \text{ is a state in }\Mo_1\} = 
\{\theTheory[\Mo_2]{t}{\PL} \mid t \text{ is a state in }\Mo_2\}.
$$
\EndDefinition

We obtain from Proposition~\ref{pres-val}

\BeginProposition{pres-val-mod}
Let $\Mo_1$ and $\Mo_2$ be models and $f: \Mo_1\to\Mo_2$ be a model morphism. Then
\begin{equation}
\label{pres-val-mod-f}
\Mo_1, s \models \phi \Longleftrightarrow \Mo_2, f(s) \models \phi
\end{equation} 
for each state $s$ of $\Mo_1$ and each formula in $\PL$. 
\EndProposition

\BeginProof
The statement is may be reformulated as
$
\Gilt_{\Mo_1} = \InvBild{f}{\Gilt_{\Mo_2}}.
$
We argue by induction on $\phi$. The equivalence in~$(\ref{pres-val-mod-f})$ is true for $\phi = \top$ and for atomic propositions by the definition of a morphism. If it is true for $\phi_1$ and for $\phi_2$, then it is also true for $\phi_1\wedge\phi_2$. 

We do an induction on program $\pi$ in formula $\md{\pi}{q}\phi$, assuming that the equivalence~$(\ref{pres-val-mod-f})$ holds for $\phi$.  If $\pi = \varrho_1;\dots;\varrho_n \in \wrd{\Ur}$, the assertion follows from Lemma~\ref{SR-isNat}, for $\pi = \pi_1\cup\pi_2$ and for $\pi = \pi_1^*$ the assertion follows from the fact that $\Phi$ and $\Psi$ are natural transformations. 
\EndProof

\medskip

Because morphisms for models and for their underlying Kripke models are the same, we obtain immediately

\BeginCorollary{same-expr}
Let $\Mo_1$ and $\Mo_2$ be models with underlying Kripke models $\K_1$ resp. $\K_2$, then
\begin{enumerate}[a.]
\item $\Mo_1$ and $\Mo_2$ are behaviorally equivalent iff $\K_1$ and $\K_2$ are behaviorally equivalent.
\item $\Mo_1$ and $\Mo_2$ are bisimilar iff $\K_1$ and $\K_2$ are bisimilar. \QED
\end{enumerate}
\EndCorollary

The construction of a model onto which logically equivalent models can be mapped requires some technical preparations, which we now turn to. 

\subsection{Factoring}
\label{sec:factoring}

The factor construction for the investigation of logical equivalence follows basically~\cite{Schubert-ISDT} and~\cite[Section 2.6.2]{EED-CoalgLogic-Book}; this construction cannot be used for the present purpose as it stands, because some small but not unimportant changes have to be made. Hence we construct factors fairly explicitly for the reader's convenience, pointing out differences as we go.

Preparing for the construction, we recall the important $\pi$-$\lambda$-Theorem from the theory of Borel sets~\cite[Theorem 1.3.1]{EED-CoalgLogic-Book}.

\BeginProposition{pi-lambda}
Let $\mathcal{A}$ be a family of subsets of a set $X$ that is closed under finite intersections. Then $\sigma(\mathcal{A})$ is the smallest family of subsets containing $\mathcal{A}$ which is closed under complementation and countable disjoint unions. 
In particular, if the measures $\mu_1, \mu_2\in\SubProb{\sigma(\mathcal{A})}$ coincide on $\mathcal{A}$, then they are equal on $\sigma(\mathcal{A})$.\QED
\EndProposition

This yields a proof strategy for the identification of $\sigma$-algebras in the construction to follow. It goes like this. In order to establish a property for all measurable sets, we will single out those sets for which the property holds and show that these sets form a generator which is closed under finite intersections. Then we will conclude through Proposition~\ref{pi-lambda} that the property holds for each set in the $\sigma$-algebra.

The following simple statement will be technically helpful as well.

\BeginLemma{invariant}
Let $f: M\to N$ be a map, and assume that $A \subseteq M$ is $f$-invariant (i.e., $a \in A$, $f(a) = f(a')$ together imply $a'\in A$). Then $\InvBild{f}{\Bild{f}{A}} = A$. If $B$ is also $f$-invariant, then $\Bild{f}{A \cap B} = \Bild{f}{A}\cap\Bild{f}{B}$. \QED
\EndLemma

Fix a model $\Mo = (\K, \Phi, \Psi)$ for the moment. Define on the state space $S$ of $\Mo$ the equivalence relation 
\begin{equation*}
\eEquiv{s}{s'}\text{ iff } \theTheory[\Mo]{s}{\PL} = \theTheory[\Mo]{s'}{\PL}.
\end{equation*}
Thus $\eEquiv{s}{s'}$ iff the state $s$ and $s'$ satisfy exactly the same PDL formulas. Define on $S$ the set $\mathcal{E}_{\text{PDL}}$ of extensions of formulas through 
\begin{equation*}
\mathcal{E}_{\text{PDL}} := \{\gilt\phi \mid \phi\text{ is a PDL formula}\}.
\end{equation*} 
Note that $\mathcal{E}_{\text{PDL}}\subseteq\Borel{S}$ is closed under finite intersections, because the logic is closed under finite conjunctions. Make the factor space $\efaktor{S}$ a measurable space by defining the $\sigma$-algebra  
\begin{equation*}
\Borel{\efaktor{S}} := \sigma(\{A\subseteq\efaktor{S} \mid \InvBild{\fMap{\ePDL}}{A} \in \mathcal{E}_{\text{PDL}} \}).
\end{equation*}
The $\sigma$-algebra is generated by the images of the formulas' extensions:

\BeginLemma{exte-gener}
The set 
$
\mathcal{A} := \{\Bild{\fMap{\ePDL}}{\gilt\phi} \mid  \phi\text{ is a PDL formula}\}
$
is a generator of $\Borel{\efaktor{S}} $ which is closed under finite intersections. If there are countably many PDL-formulas, then $\Borel{\efaktor{S}}$ is countably generated. 
\EndLemma

\BeginProof
Each extension is $\fMap{\ePDL}$-invariant by construction, the logic is closed under conjunctions, thus $\mathcal{A}$ is closed under finite intersections by Lemma~\ref{invariant}. It follows also that 
$
\gilt\phi = \InvBild{\fMap{\ePDL}}{\Bild{\fMap{\ePDL}}{\gilt\phi}},
$
thus $\mathcal{A} \subseteq \Borel{\efaktor{S}}$. Now, if $\InvBild{\fMap{\ePDL}}{A} \in \mathcal{E}_{\text{PDL}}$, then we find some PDL-formula $\phi$ with 
$
\gilt\phi = \InvBild{\fMap{\ePDL}}{A},
$
so that 
$
A = \Bild{\fMap{\ePDL}}{\gilt\phi},
$
because $\fMap{\ePDL}$ is onto. This implies $\Borel{\efaktor{S}} \subseteq \sigma(\mathcal{A})$. 

Plainly, if there are countably many PDL-formulas, then $\mathcal{A}$ is countable. 
\EndProof

\BeginCorollary{meas-fact-map}
$\fMap{\ePDL}: S \to \efaktor{S}$ is measurable.
\EndCorollary

\BeginProof
Put
$
\mathcal{D} := \{A \in \Borel{\efaktor{S}} \mid \InvBild{\fMap{\ePDL}}{A} \in \Borel{S}\},
$
then $\mathcal{D}$ is plainly closed under complementation and countable disjoint unions. We obtain from Lemma~\ref{ext-meas} and from $\gilt\phi = \InvBild{\fMap{\ePDL}}{\Bild{\fMap{\ePDL}}{\gilt\phi}}$ that $\Bild{\fMap{\ePDL}}{\gilt\phi}\in \mathcal{D}$ for each formula $\phi$, so it follows from Lemma~\ref{exte-gener} that $\mathcal{D} = \Borel{\efaktor{S}}$, from which the assertion follows. 
\EndProof

This observation permits the construction of a stochastic relation $k_\varrho: \efaktor{S}\Trans \efaktor{S}$ for each $\varrho\in\Ur$. One first notes that $\eEquiv{s}{s'}$ implies 
$
K_\varrho(s)(\gilt\phi) = K_\varrho(s')(\gilt\phi)
$
for each PDL-formula $\phi$. In fact, if, say,
$
K_\varrho(s)(\gilt\phi) < K_\varrho(s')(\gilt\phi), 
$
then we can find $q$ rational with 
$
K_\varrho(s)(\gilt\phi) < q \leq K_\varrho(s')(\gilt\phi), 
$
so that $\Mo, s \models \md{\varrho}{q} \phi$, but $\Mo, s' \not\models \md{\varrho}{q} \phi$, contradicting $\eEquiv{s}{s'}$. Consequently, $s \mapsto K_\varrho(s)(\gilt\phi)$ is constant on each $\ePDL$-class, so that 
\begin{equation*}
k_\varrho(\eklasse{s})(A) := K_\varrho(s)(\InvBild{\fMap{\ePDL}}{A}) 
\end{equation*} 
is well defined on $\efaktor{S}$ whenever $A\in\Borel{\efaktor{S}}$. Is is clear that $k_\varrho(\eklasse{s})\in\SubProb{\efaktor{S}}$, so that measurability needs to be established.

\BeginProposition{is-stoch-rel-fakt}
$k_\varrho: \efaktor{S}\Trans \efaktor{S}$ is a stochastic relation for each $\varrho\in\Ur$. 
\EndProposition

\BeginProof
Put
$
\mathcal{D} := \{A \in \Borel{\efaktor{S}} \mid v \mapsto k_\varrho(s)(A) \text{ is $\Borel{\efaktor{S}}$-measurable}\}.
$
Then evidently $\mathcal{D}$ is closed under complementation and under countable disjoint unions. Moreover, $\Bild{\fMap{\ePDL}}{\gilt\phi}\in \mathcal{D}$ for each formula $\phi$ by Lemma~\ref{exte-gener}. Because
\begin{equation*}
\{v \mid k_\varrho(v)(\Bild{\fMap{\ePDL}}{\gilt\phi}) < q\}
=
\Bild{\fMap{\ePDL}}{\gilt{\md{\varrho}{q}\phi}}
\in
\Borel{\efaktor{S}}
\end{equation*} 
we may apply Lemma~\ref{exte-gener} again, we see that 
$
\mathcal{D} = \Borel{\efaktor{S}}.
$
\EndProof

Taking $\phi = \top$, we obtain in particular from the argument above that
\begin{equation}
\label{equal-weight}
\eEquiv{s}{s'} \text{ implies }\forall\varrho\in\Ur: K_\varrho(s)(S) = K_\varrho(s')(S).
\end{equation}

\medskip

Now define the Kripke model 
\begin{equation*}
\efaktor{\K} := (\efaktor{S}, (k_\varrho)_{\varrho\in\Ur}, V_\ePDL)
\end{equation*}
with 
$
V_\ePDL := \{\Bild{\fMap{\ePDL}}{V(p)} \mid p\in\P\}
$
as the valuations for the atomic propositions. It may be noted that the equivalence relation has been defined through a model, but that we define the Kripke model now on its classes. The following observation is immediate 

\BeginLemma{is-morph-faktor}
$\fMap{\ePDL}: \K \to \efaktor{\K}$ is a morphism for Kripke models. 
\QED
\EndLemma

Define for the logically equivalent  models $\Mo_1$ and $\Mo_2$ with underlying Kripke models $\K_1$ and $\K_2$ over state spaces $S_1$ resp. $S_2$ the map $\kappa$ as follows.
\begin{equation*}
\kappa: 
\begin{cases}
\efaktor{S_1} & \to \efaktor{S_2} \\
\eklasse{s_1} & \mapsto \eklasse{s_2}\text{ iff } \theTheory[\Mo_1]{s_1}{\PL} = \theTheory[\Mo_2]{s_2}{\PL}.
\end{cases}
\end{equation*}

On account of logical equivalence, $\kappa$ is a bijection, but we can say even more. 

\BeginProposition{this-is-isom}
$\kappa: \efaktor{\K_1}\to\efaktor{\K_2}$ is an isomorphism.
\EndProposition

\BeginProof
1.
We show first that $\kappa: \efaktor{S_1} \to \efaktor{S_2}$ is measurable. In fact, let 
\begin{equation*}
\mathcal{D} := \{A \in \Borel{\efaktor{S_2}} \mid \InvBild{\kappa}{A}\in\Borel{\efaktor{S_1}}\},
\end{equation*}
then is is by Proposition~\ref{pi-lambda} and Lemma~\ref{exte-gener} enough to show that $\Bild{\fMap{\ePDL}}{\Gilt[\phi]_{\Mo_2}}\in\mathcal{D}$ for each PDL formula $\phi$. This follows from
\begin{equation*}
\InvBild{\kappa}{\Bild{\fMap{\ePDL}}{\Gilt[\phi]_{\Mo_2}}} = \Bild{\fMap{\ePDL}}{\Gilt[\phi]_{\Mo_1}}.
\end{equation*}
This implies measurability, and the equation
\begin{equation*}
\Bild{\kappa}{\Bild{\fMap{\ePDL}}{\Gilt[\phi]_{\Mo_1}}} = \Bild{\fMap{\ePDL}}{\Gilt[\phi]_{\Mo_2}}.
\end{equation*}
shows that $\kappa^{-1}$ is measurable as well. 

2.
Observe that we have 
\begin{align*}
k_{1, \varrho}(\eklasse{s_1})(\Bild{\fMap{\ePDL}}{\Gilt[\phi]_{\Mo_1}})
& =
K_{1, \varrho}(s_1)(\Gilt[\phi]_{\Mo_1}) \\
&\stackrel{(*)}{=}
K_{2, \varrho}(s_2)(\Gilt[\phi]_{\Mo_2}) \\
&=
k_{2, \varrho}(\eklasse{s_2})(\Bild{\fMap{\ePDL}}{\Gilt[\phi]_{\Mo_2}})
\end{align*}
for each $\varrho\in\Ur$ and $s_1, s_2$ with $\kappa(\eklasse{s_1}) = \eklasse{s_2}$ and for each formula $\phi$ (we argue in Equation~$(*)$ as in the proof of Corollary~\ref{meas-fact-map}). Because
\begin{equation*}
\mathcal{D} := \{A \in \Borel{\efaktor{S_2}} \mid k_{2, \varrho}(\kappa(\eklasse{s_1}))(A) 
= k_{1, \varrho}(\eklasse{s_1})(\InvBild{\kappa}{A})\}
\end{equation*}
is by~(\ref{equal-weight}) closed under complementation and countable disjoint unions, and since it contains all sets $\Bild{\fMap{\ePDL}}{\Gilt[\phi]_{\Mo_2}}$ by the argument above it equals $\Borel{\efaktor{S_2}}$ by Lemma~\ref{invariant} and by  Proposition~\ref{pi-lambda}. A very similar argument applies to $\kappa^{-1}$. 
\EndProof

These constructions can be carried out in general measurable spaces and do not need the requirement of separability, which will enter the argument in a moment.

\subsection{Logical Equivalence}
\label{sec:log-equiv}

%
%
%
%

This, then, is a characterization of logical vs. behavioral equivalence.

\BeginProposition{log-vs-behavioral-equiv}
Let $\Mo_1$ and $\Mo_2$ be models, and consider these statements.
\begin{enumerate}[a.]
\item\label{log-vs-behavioral-equiv-a} $\Mo_1$ and $\Mo_2$ are behaviorally equivalent.
\item\label{log-vs-behavioral-equiv-b} $\Mo_1$ and $\Mo_2$ are logically equivalent. 
\end{enumerate}
Then 
\begin{enumerate}[i.]
\item\label{log-vs-behavioral-equiv-i} 
$\ref{log-vs-behavioral-equiv-a}.\Rightarrow\ref{log-vs-behavioral-equiv-b}.$
\item\label{log-vs-behavioral-equiv-ii} 
If 
the set $\Ur$ of primitive programs and $\P$ of atomic propositions are countable, then $\ref{log-vs-behavioral-equiv-b}.\Rightarrow\ref{log-vs-behavioral-equiv-a}.$
\end{enumerate}
\EndProposition

\BeginProof
1.
Part$~\ref{log-vs-behavioral-equiv-i}.$ follows immediately from Proposition~\ref{pres-val-mod}, so part$~\ref{log-vs-behavioral-equiv-ii}.$ remains to be established.

2.
Let $\K_1$ and $\K_2$ be the Kripke models underlying $\Mo_1$ resp. $\Mo_2$. Construct models $\efaktor{\K_1}$ and $\efaktor{\K_2}$ and the isomorphism $\kappa:\efaktor{\K_1}\to\efaktor{\K_2}$ as in Proposition~\ref{this-is-isom}, then the state spaces of these models are separable according to Lemma~\ref{exte-gener}. 

Complete $\efaktor{\K_1}$ according to Proposition~\ref{extension-to-compl}, then we have the morphisms 
\begin{equation*}
\xymatrix{
\K_1\ar[r]^{\fMap{\ePDL}}&\overline{\efaktor{\K_1}}&&\K_2\ar[ll]_{\kappa^{-1}\circ\fMap{\ePDL}},
}
\end{equation*}
because both $\K_1$ and $\K_2$ are defined over complete spaces, again by  Proposition~\ref{extension-to-compl}. This is so because the factor map $\fMap{\ePDL}: S_1\to\efaktor{S_1}$ is also a measurable map $S_1\to\overline{\efaktor{S_1}}$. Hence $\fMap{\ePDL}: \K_1\to\efaktor{\K_1}$ extends to a morphism $\fMap{\ePDL}: \K_1\to\overline{\efaktor{\K_1}}$. A similar argument applies to $\K_2$. 

Now define $\Mo_0 := (\overline{\efaktor{\K_1}}, \Phi, \Psi)$, then $\fMap{\ePDL}: \Mo_1\to\Mo_0$ and $\kappa^{-1}\circ\fMap{\ePDL}: \Mo_2\to\Mo_0$ are the desired morphisms.
\EndProof

\section{Generalized Models}
\label{sec:g-models}

The state space of a model is assumed to be a universally complete measurable space. We relax this a bit by introducing generalized models. This is necessary in order to get a firmer grip on state spaces that are Polish, as will be argued below. 

\BeginDefinition{def-g-model}
$\No = (\K, \Phi, \Psi)$ is called an \emph{generalized model (g-model)} iff $\K$ is a Kripke model over a general measurable space; the natural transformations $\Phi: \BQ\times\BQ \nPfeil \BQ$ and $\Psi: (\BQ)^{\Nat_0} \nPfeil \BQ$ have the same properties as in Definition~\ref{Mod-PDL}. A \emph{morphism} $\No_1\to\No_2$ is a morphism for the underlying Kripke models $\K_1\to\K_2$.
\EndDefinition

Behavioral equivalence can be defined for g-models through morphisms exactly as in Definition~\ref{expr-mod}. It is, however, difficult to discuss logical equivalence, because the validity of formulas cannot be described without information about the measurable structure of the validity sets. This is so since $K_\varrho: S \Trans S$ might not be extendable to $\overline{K_\varrho}: \overline{S}\Trans \overline{S}$ in general, i.e., without additional assumptions. 

Call a Kripke model \emph{separable} iff its state space is countably generated, call accordingly an g-model \emph{separable} iff the underlying Kripke model is separable. For $\No$ separable we can construct a model $\overline{\No} = (\overline{\K}, \Phi, \Psi)$ by completion, where 
$
\overline{\K} = (\overline{S}, (\overline{K_\varrho})_{\varrho\in\Ur}, V)
$
is the completion of Kripke model $\K$. Thus we may call separable g-models $\No_1$ and $\No_2$ \emph{logically equivalent} iff their  completions $\overline{\No_1}$ and $\overline{\No_2}$ are logically equivalent. 

Assume that Kripke model $\K$ is separable. Then the inclusion $\overline{\K}\to\K$ is a morphism, hence
\begin{equation}
\label{theory-compl}
\theTheory[\K]{s}{\PM} = \theTheory[\overline{\K}]{s}{\PM}
\end{equation}
for each state $s$ of $\K$ by Proposition~\ref{pres-val}. This implies that two separable Kripke models are HM-equivalent iff their completions are HM-equivalent. 

We obtain
\BeginProposition{char-g-models}
Let $\No_1$ and $\No_2$ be separable g-models with underlying Kripke models $\K_1$ and $\K_2$. Consider 
\begin{enumerate}[a.]
\item \label{char-g-models-a} $\No_1$ and $\No_2$ are behaviorally equivalent.
\item \label{char-g-models-b} $\No_1$ and $\No_2$ are logically equivalent.
\item \label{char-g-models-c} $\K_1$ and $\K_2$ are behaviorally equivalent.
\item \label{char-g-models-d} $\K_1$ and $\K_2$ are HM-equivalent.
\item \label{char-g-models-e} $\overline{\K_1}$ and $\overline{\K_2}$ are HM-equivalent.
\end{enumerate}
Then
\begin{enumerate}[i.]
\item\label{char-g-models-i} $\ref{char-g-models-a}.\Leftrightarrow\ref{char-g-models-c}. \Leftrightarrow\ref{char-g-models-d}. \Leftrightarrow\ref{char-g-models-e}.$
\item\label{char-g-models-ii}  $\ref{char-g-models-a}. \Rightarrow\ref{char-g-models-b}.$
\end{enumerate}
\EndProposition

\BeginProof
1.
The equivalence $\ref{char-g-models-c}. \Leftrightarrow\ref{char-g-models-d}. \Leftrightarrow\ref{char-g-models-e}.$ is the first part of Theorem~\ref{main-expr} together with the observation (\ref{theory-compl}), the equivalence $\ref{char-g-models-a}.\Leftrightarrow\ref{char-g-models-c}.$ is trivial. This establishes part~$\ref{char-g-models-i}$. 

2.
If $f: \No_1\to\No_2$ is a morphism for g-models, then $f: \overline{\No_1}\to\overline{\No_2}$ is a model morphism by virtue of Proposition~\ref{extension-to-compl}. Thus part~$\ref{char-g-models-ii}.$ follows from Proposition~\ref{pres-val-mod}.
\EndProof

If we know that the separable g-models $\No_1$ and $\No_2$ are logically equivalent, and that both $\Ur$ and $\P$ are countable, then we may conclude from part~$\ref{log-vs-behavioral-equiv-ii}.$ of Proposition~\ref{log-vs-behavioral-equiv} that we can find a model $\Mo$ and surjective morphisms 
$
\overline{\No_1}\stackrel{g_1}{\longleftarrow}\Mo\stackrel{g_2}{\longrightarrow}\overline{\No_2}.
$
Tracing the construction, we even know that model $\Mo$ is the completion of a separable g-model. But there is no reason to assume that the inverse images of the morphisms $g_1$ and $g_2$ map Borel sets to Borel sets (rather than  Borel sets to universal Borel sets). 

Thus for the time being the question remains open whether logically equivalent models are behaviorally equivalent as well. 


The existence of a mediating model is dependent on topological assumptions, because ---~by the standard construction~--- a mediating model is constructed from a semi-pullback, the existence of which requires an analytic or a Standard Borel space. It is mandatory to discuss g-models in this case, because as a rule Standard Borel spaces are not complete, provided they are not countable. This can be seen as follows. Let $X$ be an uncountable Standard Borel space, then there exists an analytic set $A \subseteq X$ which is not a Borel set~\cite[Theorem 4.1.5]{Srivastava}. $A$ can be obtained through the Souslin operation  as
\begin{equation*}
A = \bigcup_{\alpha\in\Nat^\Nat}\bigcap_{n\in \Nat} F_{\alpha|n}
\end{equation*}
with a family $\{F_v\mid v \in \wrd{\Nat}\}$  of closed sets by~\cite[Theorem 4.1.13]{Srivastava}. If the measurable space $X$ would be complete, it would be closed under the Souslin operation by~\cite[Proposition 3.5.22]{Srivastava}, hence $A$ would be a Borel set, contrary to the assumption. 

We need some preparations. Let $S$ be a Standard Borel space. Call an equivalence relation $\simeq$ on $S$ \emph{countably generated} (or \emph{smooth}) iff there exists a sequence $\Folge{B}\subseteq \Borel{S}$ which defines the relation, i.e., 
\begin{equation*}
\isEquiv{s}{s'}{\simeq} \Longleftrightarrow \forall n \in \Nat: \bigl[s \in B_n \Leftrightarrow s' \in B_n\bigr].
\end{equation*} 
A set $B \subseteq S$ is called $\simeq$-invariant iff $B$ is the union of $\simeq$-classes, equivalently, iff $b\in B$ and $\isEquiv{b}{b'}{\simeq}$ together imply $b'\in B$ (hence $B$ is $\fMap{\simeq}$-invariant, see Lemma~\ref{invariant}). Relation $\simeq$ defines a $\sigma$-algebra $\mathcal{A}_\simeq \subseteq \Borel{S}$ through its invariant Borel sets, i.e., 
\begin{equation*}
\mathcal{A}_\simeq := \sigma(\{B\in\Borel{S}\mid B\text{ is $\simeq$-invariant}\}).
\end{equation*}
This construction has been studied quite extensively in the context of stochastic relations. Vice versa, this $\sigma$-algebra determines the equivalence relation uniquely~\cite{EED-CoalgLogic-Book}:

\BeginLemma{uniq-equiv-rel}
Let $S$ be a Standard Borel space with smooth equivalence relations $\simeq_1$ and $\simeq_2$. If $\mathcal{A}_{\simeq_1}= \mathcal{A}_{\simeq_2}$, then $\simeq_1 = \simeq_2$. \QED
\EndLemma

Fix a model $\Mo$ with underlying Kripke model $\K$, and assume that both $\Ur$ and $\P$ are countable. Consider these sets of formulas:
\begin{align*}
X &:= \{\md{\varrho_1}{q_1}\dots\md{\varrho_n}{q_n} p\mid p \in\P, \varrho_1, \dots, \varrho_n\in\Ur, q_1, \dots, q_n\in\Rational, n \in \Nat\}\\
Y &:= \{\langle\varrho_1\rangle_{q_1}\dots\langle\varrho_n\rangle_{q_n} p\mid p \in\P, \varrho_1, \dots, \varrho_n\in\Ur, q_1, \dots, q_n\in\Rational, n \in \Nat\}\\
Z & := \{\varphi \mid \varphi \text{ is a $\PL$-formula}\}
\end{align*}
The sets $X$ and $Y$ are countable, since $\Ur$ and $\P$ are. The formulas helping to define $X$ could be called the \emph{single-step formulas} in $\PL$: execute simple program $\varrho_n$, check whether its result on atomic sentence $p$ is below $q_n$, then execute simple program $\varrho_{n-1}$ on the corresponding states, check whether the result is below $q_{n-1}$ etc.Let $\simeq_X$  be the equivalence relations generated by the validity sets 
$
\{\gilt{\phi} \mid \phi \in X\}
$
with $\sigma$-algebras $\mathcal{A}_X$  of invariant sets, similarly for $\simeq_Y$ with $\mathcal{A}_Y$ and  for $\simeq_Z$ with $\mathcal{A}_Z$. 

This observation is obvious, because all formulas from $Z$ are generated from the formulas from $Y$ by finitary operations.
\BeginLemma{trivial}
$\mathcal{A}_Y = \mathcal{A}_Z$.
\QED
\EndLemma

Throughout the rest of the paper, we make in view of Lemma~\ref{strange-formulas} the \textbf{assumption that all Kripke models $(S, (K_\varrho)_{\varrho\in\Ur}), V)$ are strictly probabilistic}, i.e., that
\begin{equation}
\label{prob-ass}
\forall\varrho\in\Ur\forall s \in S: K_\varrho(s)(S) = 1
\end{equation}
holds. 

\BeginLemma{equal-sigma-algebras}
$\mathcal{A}_X = \mathcal{A}_Y$.
\EndLemma

\BeginProof
We infer from Corollary~\ref{simple-expr} that  $\kgilt{\psi}$ is expressible through sets from $\mathcal{A}_X$ for each $\psi\in Y$, thus $\mathcal{A}_X = \mathcal{A}_Y$. Starting from Equation~$(\ref{relate-uno})$, a similar representation of  $\gilt{\phi}$ for $\phi \in X$ through sets from $\mathcal{A}_Y$, yielding the other inclusion.
\EndProof

This has as an immediate consequence

\BeginCorollary{equiv-for-sets}
These statements are equivalent for states $s, s'$ in an g-model $\No$ with underlying Kripke model $\K$.
\begin{enumerate}[a.]
\item \label{equiv-for-sets-a} $\No, s \models \varphi \Leftrightarrow \No, s' \models \varphi$ for all single-step formulas $\phi$, i.e., all $\PM$-formulas $\varphi$ of the shape 
$
\md{\varrho_1}{q_1}\dots\md{\varrho_n}{q_n} p
$
with $\varrho_1, \dots, \varrho_n\in\Ur$, $q_1, \dots, q_n\in \Rational$, $n\in \Nat$ and $p\in\P$.
\item \label{equiv-for-sets-b} $\K, s \models \psi \Leftrightarrow \K, s' \models \psi$ for all $\PL$-formulas $\psi$.
\end{enumerate}
\EndCorollary

\BeginProof
Lemma~\ref{equal-sigma-algebras}, Lemma~\ref{trivial} and Lemma~\ref{uniq-equiv-rel}.
\EndProof

\medskip

Given g-models $\No_1$ and $\No_2$ with underlying Kripke models $\K_1$ and $\K_2$ over state spaces $S_1$ resp. $S_2$, construct the g-model 
$
\No_1\oplus\No_2 := (\K_1\oplus\K_2, \Phi, \Psi),
$
see Example~\ref{summe} with embeddings $i_{S_1}$ and $i_{S_2}$. It is not difficult to see that $S_1 + S_2$ is a Standard Borel space, provided $S_1$ and $S_2$ are, that $\overline{S_1 + S_2} = \overline{S_1} + \overline{S_2}$, and, because 
$
\No_1\stackrel{i_{S_1}}{\longrightarrow}\No_1\oplus\No_2\stackrel{i_{S_2}}{\longleftarrow}\No_2
$
are morphisms,
\begin{align*}
\overline{\No_1}, s_1 \models \phi & \Leftrightarrow \overline{\No_1\oplus\No_2}, i_{S_1}(s_1) \models \phi\\
\overline{\No_2}, s_2 \models \phi & \Leftrightarrow \overline{\No_1\oplus\No_2}, i_{S_2}(s_2) \models \phi
\end{align*} 
for all $\PM$-formulas $\phi$. 

We finally obtain for generalized models

\BeginProposition{bisim-incompl}
Let $\No_1$ and $\No_2$ be generalized models with Standard Borel state spaces, and assume that both $\Ur$ and $\P$ are countable. These statements are equivalent.
\begin{enumerate}[a.]
\item\label{bisim-incompl-a} $\No_1$ and $\No_2$ are logically equivalent.
\item\label{bisim-incompl-b} $\No_1$ and $\No_2$ are behaviorally equivalent.
\item\label{bisim-incompl-c} $\No_1$ and $\No_2$ are bisimilar.
\end{enumerate}
\EndProposition

\BeginProof
0.
Because Standard Borel spaces are based on Polish spaces which in turn have a countable base for their topology, the g-models under consideration are countably based.

1.
\labelImpl{bisim-incompl-b}{bisim-incompl-c}: Assume that $\No_1$ and $\No_2$ are logically equivalent. Let $\K_1$ and $\K_2$ be the underlying Kripke models with state spaces $S_1$ and $S_2$ and valuations $V_1$ resp. $V_2$. We claim that $\K_1$ and $\K_2$ are HM-equivalent. Given $s\in S_1$ there exists $s'\in S_2$ with
$
\theTheory[\overline{\No_1}]{s}{\PM} = \theTheory[\overline{\No_2}]{s'}{\PM}
$
so that  
\begin{equation*}
\label{valid-for-N}
\overline{\No_1}, s \models \phi \Leftrightarrow \overline{\No_2}, s' \models \phi
\end{equation*}
holds for all $\PM$-formulas $\phi$, thus
\begin{equation*}
\label{valid-for-N+N}
\overline{\No_1\oplus\No_2}, i_{S_1}(s) \models \phi \Leftrightarrow \overline{\No_1\oplus\No_2}, i_{S_2}(s') \models \phi.
\end{equation*}
This holds in particular for all formulas of the syntactic shape given in part~$\ref{equiv-for-sets-a}.$ of Corollary~\ref{equiv-for-sets}, from which we infer that
\begin{equation*}
\label{valid-for-K+K}
\K_1\oplus\K_2, i_{S_1}(s) \models \psi \Leftrightarrow \K_1\oplus\K_2, i_{S_2}(s') \models \psi
\end{equation*}
holds for all $\PL$-formulas $\psi$, thus
\begin{equation*}
\label{valid-for-K}
\K_1, s \models \psi \Leftrightarrow \K_2, s' \models \psi
\end{equation*}
is inferred for all $\PL$-formulas $\psi$. Hence $\K_1$ and $\K_2$ are HM-equivalent by Proposition~\ref{char-g-models}, so that $\No_1$ and $\No_2$ are bisimilar by Corollary~\ref{same-expr}. 
\EndProof

\section{Conclusion and Further Work}
\label{sec:conclusion}

We investigate propositional dynamic logics (PDL) with a view towards a coalgebraic interpretation. This logic is technically a bit more challenging than the usual modal logics because its modalities do not always correspond to the interpreting relations in a Kripke model. Hence these relations have to be provided, which is straightforward for non-deterministic Kripke models, but turns out to be somewhat involved in the case of their stochastic counterpart. This is so since there are no natural counterparts to the program constructs in the set of stochastic relations. We observe also that interpreting PDL makes some informal assumptions on the programs' semantics like associativity over the basic operations or some sort of distributivity of program composition and the nondeterministic choice. 

In order to prepare the ground for a coalgebraic interpretation we have a closer look at the programs; they are perceived as elements of a term algebra, the primitive terms being taken from a set of primitive programs. The informal semantics is translated into a set of rewrite rules and equations; it turns out that we have to adjust the term algebra a bit when looking at the indefinite iteration of a program. Each program is shown to correspond to an irreducible one, unique up to the congruence made up from the rewriting rules and the equations. This irreducible program can easily be interpreted in a coalgebra, because we have eliminated the crucial indefinite iteration and replaced it by an operation which is easier to handle (but there is no free lunch: we pay the price for this by an operation of infinite arity). 

We specialize the coalgebraic discussion for most of the paper to  coalgebras related to the subprobability functor. They are discussed and brought into the interpretation. This is followed by the investigation of the expressivity of the corresponding models. Due to some measure-theoretic observations we have to discuss these questions with a distinct look for the details, i.e., for the particulars of the underlying state spaces. It turns out to be helpful to complete a model and to study the interplay of completion and expressivity.

Further work will include applying the present approach to game logics as proposed by Parikh~\cite{Parikh-Games1985}, see also~\cite{Pauly-Parikh}. A first step towards a coalgebraic interpretation can be found in~\cite{EED-Game-Coalg}, where in particular the notions of bisimilarity from~\cite{Parikh-Games1985,Pauly-Parikh} has been related to the one studied in coalgebras~\cite{Rutten}. 

While the present approach deals mainly with stochastic relations and the corresponding predicate liftings, the use of term rewriting can certainly be applied for defining the coalgebraic semantics of dynamic logics for other functors. 

\paragraph{Acknowledgements.} The author wants to gratefully acknowledge discussions with Chunlai Zhou, Christoph Schubert, Shashi Srivastava and H. Sabadhikari.

\end{document}